\documentclass[fleqn,usenatbib,twocolumn]{mnras}

\usepackage{newtxtext,newtxmath}

\usepackage[T1]{fontenc}
\usepackage{ae,aecompl}

\usepackage{graphicx}	
\usepackage{amsmath}	
\usepackage{amssymb}	
\usepackage{color}
\usepackage[usenames,dvipsnames,svgnames,table]{xcolor}
\hypersetup{
    citecolor = {gray},
    linkcolor = {black}
}

\newcommand{\bn}{\begin{enumerate}}
\newcommand{\en}{\end{enumerate}}
\newcommand{\bi}{\begin{itemize}}
\newcommand{\ei}{\end{itemize}}

\newcommand{\Zsun}{Z_\odot}

\newcommand{\Msun}{M_\odot}

\newcommand{\ccc}{{\rm cm}^{-3}}
\newcommand{\Htwo}{{\rm H_2}}

\title[Baseline Enrichment]{Baseline Metal Enrichment from Population III Star Formation in Cosmological Volume Simulations}

\author[Jaacks et al.]{Jason Jaacks$^1$\thanks{email: jaacks@astro.as.utexas.edu}, Robert Thompson$^{2,3}$, Steven L. Finkelstein$^1$ and Volker Bromm$^1$ \vspace{0.3cm}\\
$^1$ Department of Astronomy, The University of Texas at Austin, Austin, TX 78712\\
$^2$ Portalarium Inc., Austin, TX 78731\\
$^3$ National Center for Supercomputing Applications, University of Illinois at Urbana-Champaign, Champaign, IL 61801
}

\date{Accepted XXX. Received YYY; in original form ZZZ}

\pubyear{2017}

\begin{document}
\label{firstpage}
\pagerange{\pageref{firstpage}--\pageref{lastpage}}
\maketitle


\begin{abstract}
We utilize the hydrodynamic and N-body code {\small GIZMO} coupled with our newly developed sub-grid Population~III (Pop~III) Legacy model, designed specifically for cosmological volume simulations, to study the baseline metal enrichment from Pop~III star formation at $z>7$.  In this idealized numerical experiment, we only consider Pop~III star formation. We find that our model Pop~III star formation rate density (SFRD), which peaks at $\sim 10^{-3}\ {\rm \Msun yr^{-1} Mpc^{-1}}$ near $z\sim10$, agrees well with previous numerical studies and is consistent with the observed estimates for Pop~II SFRDs. The mean Pop~III metallicity rises smoothly from $z=25-7$, but does not reach the critical metallicity value, $Z_{\rm crit}=10^{-4}\ \Zsun$, required for the Pop~III to Pop~II transition in star formation mode until $z\simeq7$.  This suggests that, while individual halos can suppress in-situ Pop~III star formation,  the external enrichment is insufficient to globally terminate Pop~III star formation. The maximum enrichment from Pop~III star formation in star forming dark matter halos is $Z\sim10^{-2}\ \Zsun$, whereas the minimum found in externally enriched haloes is $Z\gtrsim10^{-7}\ \Zsun$.  Finally, mock observations of our simulated IGM enriched with Pop~III metals produce equivalent widths similar to observations of an extremely metal poor damped Lyman alpha (DLA) system at $z=7.04$, which is thought to be enriched by Pop~III star formation only.
\end{abstract}



\section{Introduction}
\label{sec:intro}

\begin{table*} 
\caption{Simulation parameters used in this paper. The parameter $N_{\rm p}$ is the number of gas and dark matter particles; $m_{\rm DM}$ and $m_{\rm gas}$ are the particle masses of dark matter and gas; $\epsilon,h_{\rm sml}$ are the comoving gravitational softening length/hydrodynamical smoothing length (adaptive). Both simulation runs utilize the Pop~III Legacy model (P3L), detailed in Section~\ref{subsec:legacy}, and differ only in the inclusion of the proxy for Pop~II star formation (see Section~\ref{subsubsec:P2P}).}
\begin{tabular}{cccccccc}
\hline
Run  	& Box size 	& $N_{\rm p}$ 		& $m_{\rm DM}$ 			& $m_{\rm gas}$ 		 & $\epsilon,h_{\rm sml}$  & Pop~III 	&	Pop~II\\ 
			& (Mpc $h^{-1}$) 	& (DM, Gas) 	& ($\Msun$)		& ($\Msun$) 			 & (kpc)  	&	model	&	model\\
\hline

N512L4 		& $4.0$ 	& $2 {\times} 512^{3}$ 	& $4.31{\times} 10^{4}$ 	& $9.64 {\times} 10^{3}$ & $0.45$ 	&P3L	& None \\
N512L4\_P2P 		& $4.0$ 	& $2 {\times} 512^{3}$ 	& $4.31{\times} 10^{4}$ 	& $9.64 {\times} 10^{3}$ & $0.45$  	& P3L	& P2P\\
\hline
\end{tabular} 
\label{tbl:Sim}
\end{table*}

Massive, short lived, and forming from nearly pristine (metal free) environments, the first generation of stars to form in the Universe, the so-called Population~III (Pop~III), plays a critical role in the cosmic evolutionary sequence.  Metals left behind by the death of Pop III stars in supernova (SN) explosions enrich the early interstellar/intergalactic medium (ISM/IGM) \citep[e.g.][]{Yoshida:04} to a critical metallicity \citep[$Z_{\rm crit}\sim 10^{-4}\ \Zsun$,][]{Bromm:01a}, required to engender a second generation of low-metallicity, predominantly lower mass Population~II (Pop~II) stars. The transition from a top-heavy primordial initial mass function (IMF) to a lower mass IMF \citep{Salpeter:55,Chabrier:03} marks a critical point in cosmic evolution as Pop~II stars constitute the high redshift galaxies which are the focus of current surveys (e.g. HUDF, CANDELS, BoRG, HST Frontier fields), and ultimately the old populations of galaxies we observe locally. Further, the UV ionizing photons produced during the life and death of massive Pop~III stars make an important contribution to ending the cosmic dark ages \citep[e.g.][]{Tumlinson:00,Bromm:01b,Schaerer:02,Schaerer:03}.  This renders understanding where and when Pop~III stars formed and evolved essential to understanding galaxy evolution in the first billion years of the Universe.  

Due to their short life spans and the epoch in which they dominate cosmic star formation ($z\gtrsim 15$) \citep[e.g.][]{Tornatore:07,Pallottini:15,Xu:16}, direct observations of Pop~III stars and star forming regions is beyond the capabilities of even our most powerful telescopes \citep[e.g.][]{Cai:11,Pawlik:11,Cassata:13,Sobral:15,Kehrig:15}.  We therefore must rely on indirect evidence derived from observations of lower redshift systems.  Damped Lyman alpha (DLA) and Lyman alpha (Ly$\alpha$) absorbers, as well as Lyman limit systems, can offer insight into the column densities and metallicities of the IGM near distant, faint sources.  For example, \citet{Rafelski:14}, using high resolution spectroscopy to study DLAs, found evidence that the mean metallicity of neutral gas rapidly drops at $z\sim 5$ to $\left<Z\right>\simeq10^{-2.0}\ \Zsun$.  Using similar techniques, \citet{Simcoe:12} detected a system of high column density neutral gas at $z$=7.04 (if the system is bound), estimated to have $Z<10^{-4}\ \Zsun$, which is near the critical metallicity for the Pop III to Pop II transition.   However, these systems have only been observed at $z\lesssim 8$, thus showing potential contamination from Pop~II star formation. To disentangle any such multi-component enrichment, we must therefore rely on numerical simulations, tracking in detail the contribution from Pop~III star formation to the metallicity of the early Universe.

Early analytical efforts to derive the cosmological impact of Pop~III star formation \citep{Carr:84} considered the contribution of Pop~III remnants to the ``missing matter'' (i.e. dark matter) problem, the contribution to reionization and the enrichment of the IGM.  It was determined in this work that the remnants of Pop~III stars could significantly contribute to the ``missing matter'' if their IMF was Salpeter-like, implying a significant number of stars with $M<0.1\Msun$.  It was also determined that intermediate mass Pop~III stars could be an important contributor to reionization and pre-galactic enrichment.

 \citet{Yoshida:04} concluded, based on cosmological simulations from \citet{Yoshida:03}, that a significant fraction of the IGM could be enriched to a critical value of $10^{-3.5}\ \Zsun$ by $z$=15.  It was also concluded that Pop~III star formation was a ``self-terminating'' process due to the rapid enrichment and radiative feedback produced.  This result is however based on the assumption that Pop~III stars had masses of $100-300\ \Msun$, thus ending their life as pair instability supernovae (PISNe), with metal yields $\sim 5$ times higher compared to their lower mass counterparts. Since this study, advancements in computational power have allowed for simulations with greatly increased mass and spatial resolutions leading to the discovery that, due to fragmentation, not all Pop~III stars have masses $>100\ \Msun$ \citep{Turk:09,Stacy:10}. Consequently, far fewer PISN events are predicted.  The results from \citet{Yoshida:04} can be looked at now as an upper limit for the metallicity evolution.

Utilizing smoothed particle hydrodynamics (SPH) simulations, \citet{Tornatore:07} found that, while experiencing rapid quenching by $z\sim 6$, Pop~III star formation could continue down to $z\approx 2.5$, suggesting that direct observations of systems powered by Pop~III stars could be possible.  They also concluded that only a small fraction of the baryons at $z=3$ ($\Omega_{\rm III}\sim 10^{-5}\Omega_{b}$) are enriched by Pop~III star formation only, making detection of Pop~III enriched gas regions difficult as the vast majority have been further enriched by subsequent Pop~II star formation.  While cutting edge at the time, these simulations did not fully track the primary primordial coolants ($\Htwo, {\rm HD}$) and lacked the spatial resolution to fully resolve Pop~III star forming regions. Further, as with \citet{Yoshida:04}, only PISNe were considered for metal feedback/enrichment.

A more recent study by \citet{Pallottini:15} utilized adaptive mesh refinement (AMR) simulations to study the cosmic metal enrichment by the first galaxies.  The use of AMR allows for enhanced spatial resolution in pre-selected regions of interest (i.e. high density star forming regions). This work estimates the Pop~III star formation rate density (SFRD) to peak at $\sim 10^{-3} \Msun$~yr$^{-1}$~Mpc$^{-3}$ at $z$=9, which is only a factor of ten lower than the observed SFRD, thought to be dominated by Pop~II star formation.  It was also estimated that the volume filling fraction of gas with metallicities $Z>Z_{\rm crit}$ is $\sim 10^{-4}$ at $z$=10 and $\sim 10^{-2.5}$ at $z$=6.  This estimate however did not distinguish between the contribution of metals originating from Pop~III as compared to those from Pop~II stars.

Studying the enrichment from Pop~III stars only, which is the focus of this work, we are able to determine the baseline enrichment in our simulated Universe.  Thus having an estimate for the Pop~III metal enrichment will help to provide insight into the impact, frequency, origins and evolutionary path of extremely low metallicity systems such as Cosmos Redshift 7 (CR7), an extremely luminous Lyman-$\alpha$ source that is seemingly devoid of metals \citep{Sobral:15}, and the \citet{Simcoe:12} DLA.  Theoretical studies have suggested that a direct collapse black hole (DCBH) is a more likely explanation for the \citet{Sobral:15} observations \citep[e.g.][]{Pallottini:15b,Smith:16,Agarwal:16,Agarwal:17,Pacucci:17}. Also, follow up observations of this system have also suggested that it may be more metal enriched than previously estimated \citep{Bowler:17,Matthee:17,Sobral:17}.  In either case this work will also allow us to provide a heuristic map of CR7-type objects which can be utilized by current HST wide field surveys and future {\it James Webb Space Telescope (JWST)} observations and which will probe even earlier epochs where such systems would, in theory, be more numerous \citep{Smith:17}.

In this work we detail a newly developed Pop~III star formation model in our cosmological volume simulations and utilize it to study the baseline metal enrichment legacy left by Pop~III stars at $z\geq 7$. We would like to point out to the reader, when examining these results, that we are only considering metal enrichment from Pop~III star formation. Therefore, what we present here is an idealized numerical experiment, providing an upper limit to the ability of Pop~III to enrich the early Universe. The paper is structured as follows.  In Section~\ref{sec:sim} we describe the simulations and methods used. Specifically, we provide details of our new Legacy Pop~III model in Sec~\ref{subsec:legacy}.  In Section~\ref{sec:results} and \ref{sec:sum} we present our results and conclusions.

\begin{figure*}
\begin{center}
\includegraphics[scale=0.30] {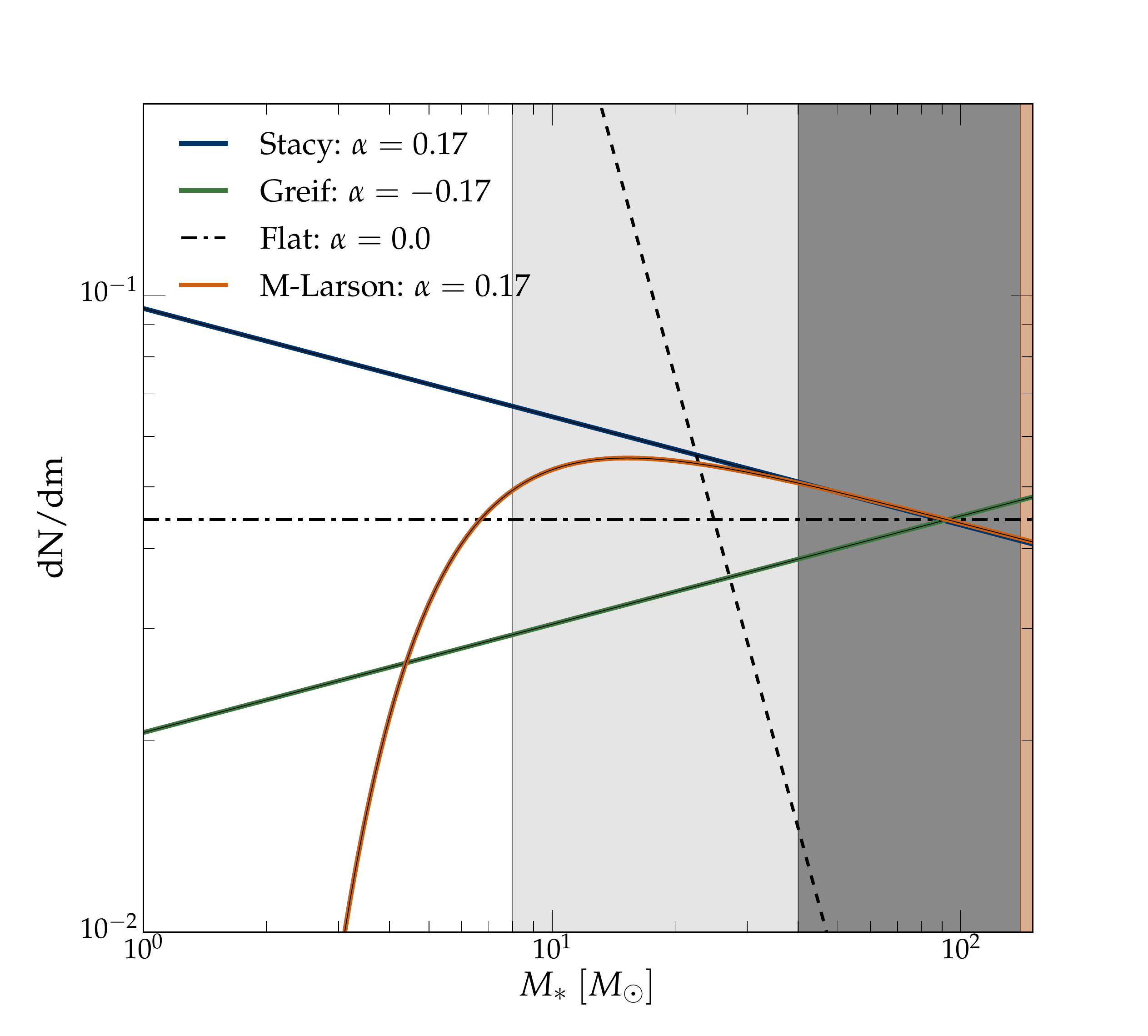}
\includegraphics[scale=0.30] {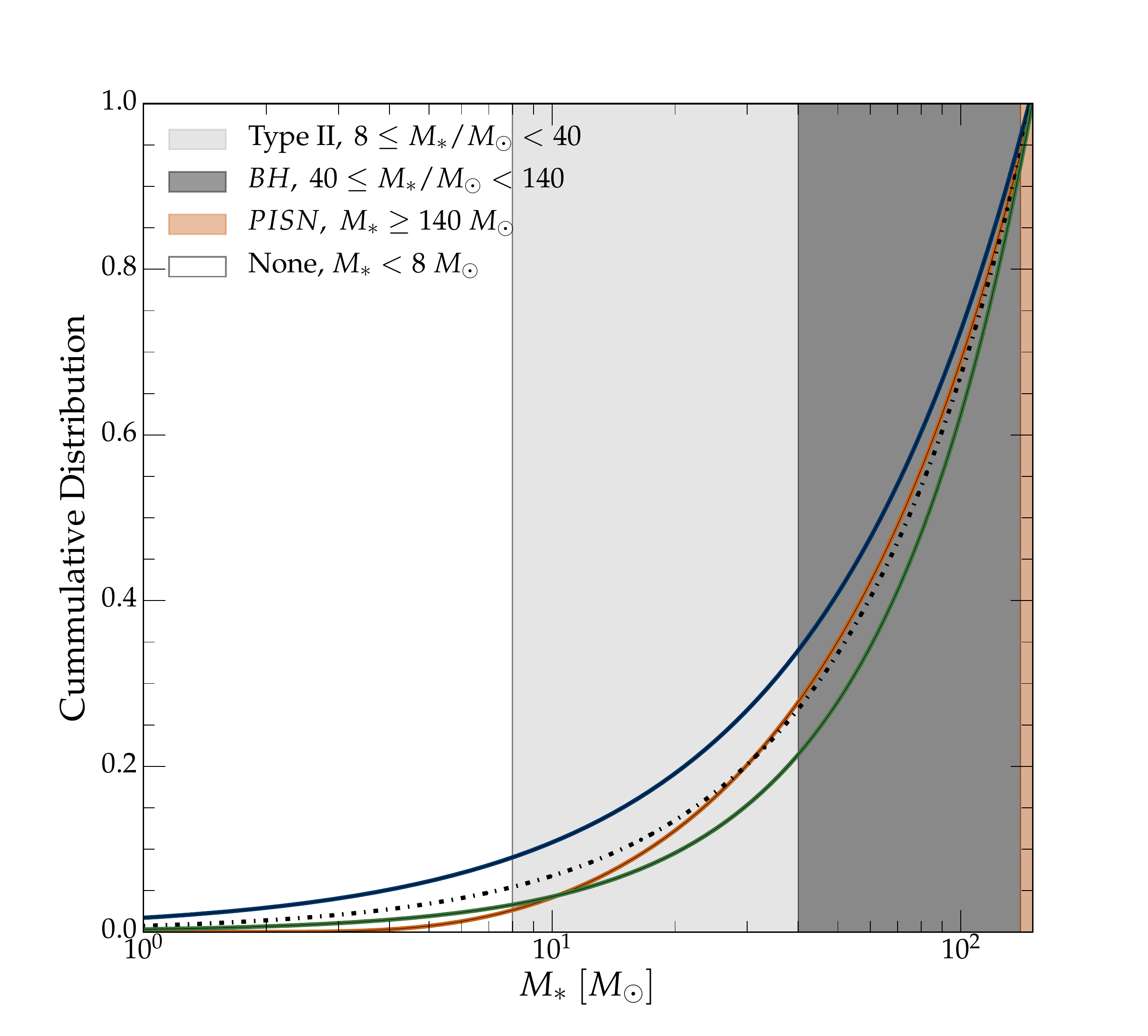}
\caption{{\it Left:} Compilation of several theoretical models for Pop~III initial mass functions which are normalized to our total stellar mass of $M_*=500\ \Msun$. Shown are three power-law IMFs with varying slopes of $\alpha=0.17,-0.17,0.0$ \citep{Stacy:13,Greif:11}, and one which is a modified \citet{Larson:98} IMF with a steeper exponential cut-off, designed to minimize low mass star formation ($\alpha=0.17, M_{\rm cut}=4.47\ \Msun$). For comparison, the dashed black line represents a power-law with slope  $\alpha=2.35$ \citep{Salpeter:55}, consistent with a local IMF.
{\it Right:} The cumulative probability distribution function for each of the IMFs shown on the left, utilized for the random draw process in our star formation algorithm. The shaded regions in both panels indicate the mass range associated with the different stellar fates (i.e. no feedback, Type~II SN, black hole, or PISNe).}
\label{fig:imf}
\end{center}
\end{figure*} 


\section{Numerical Methodology}
\label{sec:sim}
\begin{table*} 
\caption{Mean values of stellar mass, $\left< M_* \right>$, mean total number of stars, $\left< N_*\right>$, and fractional contribution to the total number from each mass range, $\left < N_i/N_* \right>$, calculated for each of the IMFs studied. We derive the averages for $10^5$ unique star forming events, drawn from each of the IMFs indicated.}
\begin{tabular}{cccccccccc}
\hline
 Run     & IMF & $\alpha$ & $\eta_*$ & $\left<M_*\right>\ [\Msun]$ & $\left<N_*\right>$ & $\left<N_{\rm TypeII}/N_*\right>$ & $\left<N_{\rm BH}/N_*\right>$ & $\left<N_{\rm PISNe}/N*\right>$ & $\left<N_{\rm low mass}/N_*\right>$\\
\hline
\vspace{0.05 mm}\\
N514L4      & Stacy 	& $0.17$  	& $0.05$  & $547.57$	& $8.12$  & $0.25$  & $0.62$   & $0.06$  & $0.07$\\
	     & Greif 	& $-0.17$  	& $0.05$  & $550.60$	& $6.76$  & $0.18$  & $0.71$   & $0.08$  & $0.03$\\
         & Flat 	& $0.0$  	& $0.05$  & $549.64$	& $7.31$  & $0.21$  & $0.67$   & $0.07$  & $0.05$\\
         & M-Larson & $0.17$  	& $0.05$  & $548.48$	& $7.49$  & $0.25$  & $0.66$   & $0.06$  & $0.03$\\
\vspace{0.05 mm}\\
\hline
\end{tabular}
\label{tbl:imf}
\end{table*}

This work takes advantage of recent advancements in hydrodynamical numerical techniques through the utilization of the public version of the new {\small GIZMO} \citep{Hopkins:15} simulation code.  {\small GIZMO} employs a Lagrangian meshless finite-mass (MFM) method for solving the equations of hydrodynamics which improves on previous generations of smoothed particle hydrodynamics (SPH) and adaptive mesh refinement (AMR) codes (see \citet{Hopkins:15} for method comparison details). In this section, we will describe our newly developed sub-grid physics modules, along with the physical parameters adopted in this work.

Our simulation volume, designed to approximately replicate a single pointing with the {\it JWST} at redshift $z\sim 10$, has a box size of $4h^{-1}$ Mpc and contains $512^3$ particles of both gas and dark matter.  We will refer to this simulation run as N514L4 throughout.  Full details of the simulation set-up can be found in Table~\ref{tbl:Sim}.  We adopt a $\Lambda$ cold dark matter ($\Lambda$CDM) cosmology, consistent with the recent {\it Planck} results: $\Omega_m=0.315$, $\Omega_\Lambda=0.685$, $\Omega_b=0.047$, $\sigma_8=0.829$, and $H_0=67.74$ \citep{Planck:16}. Our initial conditions are generated at $z=250$, using the {\small MUSIC} initial conditions generator \citep{Hahn:11}.

Primordial chemical abundances and radiative cooling rates are calculated for 12 species (H, H$^+$, H$^-$, $\Htwo$, $\Htwo^+$, He, He$^+$, He$^{2+}$, D, D$^+$ HD, e$^-$), using methods detailed in \citet{Bromm:02} and \citet{Johnson:06}, in turn based on earlier work \citep{Cen:92,Galli:98}. As $\Htwo$ and HD are the primary low-temperature coolants in primordial gas, it is critical to properly account for their formation and destruction.  Therefore, we also include $\Htwo$ photo-dissociation, and photo-detachment of ${\rm H^-,\Htwo^+}$ in our chemical network (see Section~\ref{subsub:jlw} for details).

Dark matter haloes are identified via a post processing 3D friends-of-friends (FOF) algorithm with a minimum particle requirement of $32$ and a linking length of $0.15$ times the inter-particle distance.  Gas particles and their respective properties (mass, temperature, metallicity, density, position) are then associated with each dark matter halo by searching within its virial radius. Grouping and data extraction are aided by the {\it yt} \citep{Turk:yt} and \href{https://bitbucket.org/rthompson/caesar}{Ceasar} \citep{pygr} software packages. 

\subsection{Legacy Pop~III star formation}
\label{subsec:legacy}

Formation of Pop~III stars, and the resulting supernova (SN) pollution of the surrounding medium with metals, is vital to accurately predict where and when the second generation of star formation will occur and, by extension, the first galaxies form. To model Pop~III SN explosions in full detail would require sub-parsec scale resolution  \citep[e.g.][]{Ritter:12}, which is not achievable in cosmological volume simulations.  Therefore, we implement what we term the ``Legacy`` approach, where we are concerned with modeling the final impact of an episode of Pop~III star formation, specifically the resulting metal enrichment, persistent ionization, and thermal energy, but where we do not model the feedback event itself.  

Our Legacy approach allows us to form Pop~III stellar populations from a randomly sampled input IMF after which we essentially ``paint'' fully formed Pop~III supernova remnants onto the fluid, centered on the star forming regions with physical properties calibrated with high-resolution simulations \citep[e.g.][]{Greif:07,Stacy:13}, and with select analytic solutions \citep{Sedov:59,Taylor:50}.  Each star formation event gives rise to a stellar population which is randomly drawn from a given initial mass function (IMF).  This allows each star forming region to exhibit a unique feedback signature, involving energy and nucleosynthetic yield, as stars with different masses end their lives differently (i.e. type~II SN, black hole, pair instability SN). The feedback radius of influence, metallicity, thermal energy and ionization are directly calculated for each individual stellar population.  Thereafter, the enriched gas simply advects with the local hydrodynamical flow.  This approach enables Pop~III star formation to have both time and spatial dependence, rather than the idealized uniform metallicity floor which was implemented in our previous simulations \citep{Jaacks.etal:12a,Jaacks.etal:12b,Jaacks:13,Thompson:14}.

\subsubsection{Star formation criteria}
\label{sec:sf_cri}

Pop~III star formation is triggered when a preset threshold density, $n_{\rm th}=100$~cm$^{-3}$, is reached for a gas (SPH) particle with $T\leq 10^3$ K and $Z \leq Z_{\rm crit}$. Here, the critical metallicity required to transition from Pop~III star formation to Pop II is $Z_{\rm crit}\approx10^{-4} Z_{\odot}$ \citep{Bromm:01a,Schneider:02}.  This density is adopted due to the mass/spatial resolution limitations of our simulation volumes, which are only able to resolve molecular cloud scale objects and not the individual star forming clumps contained within. The choice of  $n_{\rm th}=100$~cm$^{-3}$ as our star formation threshold is to ensure that the gas has reached a density such that cooling via molecular processes is efficient.

\subsubsection{Stellar population}
\label{sec:ssp}

Our mass resolution of $3.0\times10^5\ \Msun$, where $M_{\rm res}=m_{\rm gas} N_{\rm neighbor}$, is insufficient to simulate the formation of individual Pop~III stars. Therefore, we assume the formation of a simple stellar population (SSP), with a total mass of $M_{*,\rm Pop~III}\sim\ 500\Msun$. This mass corresponds to a primordial pre-stellar core, as found in high resolution simulations of Pop~III star formation in minihalo hosts \citep{Greif:11,Hirano:14,Stacy:16}.  We adjust the star formation efficiency parameter, $\eta_*$, for each run such that $M_{*,\rm Pop~III}=\eta_*m_{\rm gas}$.  Here $m_{\rm gas}$ is the mass of a single gas particle in a given simulation run (see Table~\ref{tbl:Sim}), implying a Pop~III star formation efficiency of 
$\eta_*\approx0.05$ (see Table~\ref{tbl:imf}).

Each star formation event is assigned a unique stellar population which is randomly drawn from a user provided initial mass function (IMF), until the sum of the individual stellar masses exceeds, or is equal to, the desired total mass of the cluster
\begin{equation}
\sum_{i=1}^{N_{*}}M_{*,i}\geq M_{*,\rm Pop~III}
\end{equation}
Here, $M_{*,i}$ is determined by a random number, uniformly drawn between $[0-1]$, which is then associated with the corresponding stellar mass via a cumulative probability distribution function, derived from the IMF (see Figure~\ref{fig:imf}).  The summation is required to reach at least the desired stellar mass to avoid biasing our stellar population towards low mass stars in our random draw.  We do not insert a ``star particle'' tracer into the simulation, implying that the information is lost once the feedback loop has completed.

\subsubsection{Initial Mass Function}
\label{sec:imf}

The true nature of the Pop~III IMF has yet to be determined. From the theory side, the field must rely on high resolution hydrodynamical simulations to study the IMF.  Unfortunately, there is no clear convergence from the various studies towards a consensus shape or slope.  Due to the lack of metals to cool the primordial gas in the very early Universe, the typical Jeans mass of proto-stellar clouds is thought to lead to higher mass stars when compared to local star formations. Estimates of typical Pop~III stellar masses are on the order of $20-100\ \Msun$, in contrast to the $<1\ \Msun$ we observe locally \citep{Salpeter:55}. Due to the uncertainty in the Pop~III IMF, we present several IMFs from the literature as an illustration the differences.

In Figure~\ref{fig:imf}, we present a compilation of Pop~III IMFs explored in this study.  Three of these have been determined by various studies using high resolution hydrodynamical simulations \citep{Stacy:13,Greif:11}, and can be characterized by a simple power-law
\begin{equation}
\Phi(M)\propto M_*^{-\alpha}
\end{equation}
with $\alpha=-0.17,0.0,0.17$.  A third option is a modification of the Larson IMF \citep{Larson:98}, which consists of an exponential cut-off to a power-law slope originally developed to exclude very low mass stars in the local IMF.  We utilize this cut-off in much the same way (i.e. reduce the probability of stars with $M_*<8\ \Msun$), and include a modification which squares the exponential terms to promote a more rapid turn-over:

\begin{equation}
\Phi(M)\propto M_*^{-\alpha} \exp\left({\frac{-M_{\rm cut}^2}{M^2_*}}\right)
\label{eq:mlarson}
\end{equation}
where $\alpha=0.17$ and $M_{\rm cut}=4.47\ \Msun$.  For contrast we also include a typical local Salpeter IMF \citep{Salpeter:55} with a slope of $\alpha=2.35$ (dashed gray line).  We normalize each IMF to the total Pop III mass ($M_{*,\rm PopIII}$) by integrating over a mass range of $1\ \Msun$ to $150\ \Msun$

\begin{equation}
M_{*,\rm PopIII} = A\int^{150}_1 \Phi(M)\ M_*\ dM_*
\label{eq:imf}
\end{equation}
where A is the normalization constant.  This mass range allows for a stellar population which can have several post main-sequence evolutionary outcomes, as follows:

\begin{eqnarray*}
\ \ \ \ \ \ \ \ \ 140\leq M/\Msun  \leq 150 	&\rightarrow& 	{\rm Pair-instability\ SN\ (PISN)} \\ 
\ \ \ \ \ \ \ \ \ 40\leq  M/\Msun     < 140 	&\rightarrow&	{\rm Black\ hole\ (BH)} \\ 
\ \ \ \ \ \ \ \ \ 8\leq   M/\Msun      < 40 	&\rightarrow&	{\rm Core\ collapse\ SN\ (Type~II)} \\
\ \ \ \ \ \ \ \ \ 		  M/\Msun       < 8		&\rightarrow&	{\rm Low\ mass\ stars}
\end{eqnarray*}

Each of which will contribute a different amount of energy and metallicity to the total budget of the subsequent SN remnant \citep{Heger:02}.  The upper mass limit is chosen to reflect recent observational constraints on the rarity of PISN events in establishing the fossil chemical abundance record in extremely metal-poor stars \citep[e.g.][]{Fraser:17}.

\begin{figure}
\begin{center}
\includegraphics[scale=0.32] {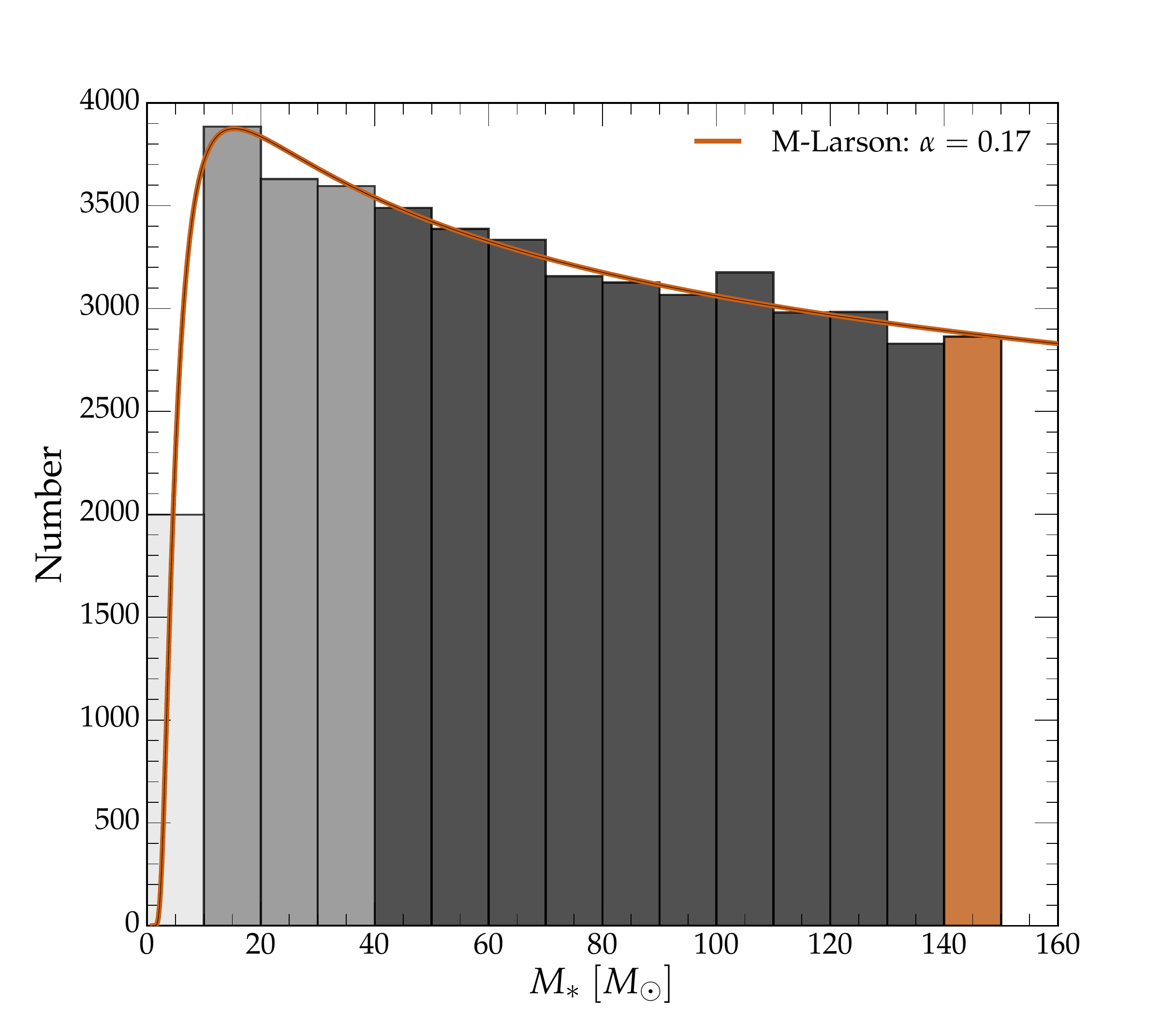}
\caption{Histogram of $10^5$ unique star formation events, randomly drawn from the M-Larson IMF ($\alpha=0.17,\ M_{\rm cut}=4.47\ \Msun$). This test demonstrates that our sampling procedure indeed reproduces the intended global IMF (solid orange line).  The shaded regions represent the same mass ranges as in Figure~\ref{fig:imf}.}
\label{fig:hist}
\end{center}
\end{figure}

Prior to implementation into {\small GIZMO}, a sample set of $10^5$ stellar populations was generated to determine the ability of our randomly drawn Legacy Pop~III SF method to reproduce a global IMF.  The results of this test are presented in Figure~\ref{fig:hist}, where the histogram gives the distribution of the global population as compared to the input IMF (solid orange line). Evidently, our sampling procedure reproduces the input IMF extremely well. We further illustrate this in Table~\ref{tbl:imf}, where we provide the mean total mass ($\left< M_* \right>$), mean total number of stars ($\left< N_*\right>$), and fractional contribution to the total number from each mass range ($\left < N_i/N_* \right>$), calculated for each of the IMFs represented by $10^5$ unique star forming events.

We note that, based on the data summarized in Table~\ref{tbl:imf} and results presented in \citet{Pallottini:15}, our results are mostly insensitive to the detailed choice of Pop~III IMF. For this reason, we here only show results for the IMF given by Equation~\ref{eq:mlarson}.

\subsection{Legacy Feedback Prescription}
\label{sec:SN_bubble}
\subsubsection{Enrichment radius}
The physics of the initial SN explosion and the subsequent interaction with the surrounding interstellar medium (ISM) can be described by well-known analytical models for each phase of the expansion. Employing the solution for each phase (i.e. free expansion, Sedov-Taylor, pressure/momentum-driven snowplow), we can robustly predict the radius of the SN shell, $R_{\rm SN}$, which can then be compared to detailed, high-resolution hydrodynamics simulations.

\begin{figure}
\begin{center}
\includegraphics[scale=0.30] {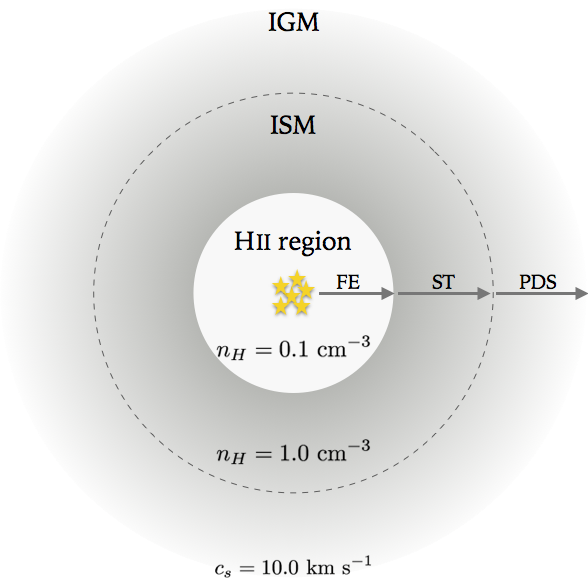}
\caption{Conceptual illustration of typical star-forming region in our model, along with physical properties assumed in each region for the post star-burst, pre-SN phase.  Physical properties are taken to be consistent with simulations of \ion{H}{ii} regions around Pop~III stars \citep{Kitayama:05}.}
\label{fig:region}
\end{center}
\end{figure}

Prior to the SN explosion, ionizing radiation and stellar winds have established a low density region surrounding the host star (see Figure~\ref{fig:region} for physical properties adopted).  Since the supernova remnant (SNR) has a much higher density than the surrounding medium it expands freely at early times.  This expansion continues until the mass swept up by the expanding SNR becomes equal to the mass of the ejecta ($M_{\rm sw}=M_{\rm ej}\approx M_{*,\rm PopIII}$).  The radius at which this equality occurs can be estimated by 
\begin{equation}
r_{\rm fe}\approx\left( \frac{3XM_{\rm ej}}{4\pi m_p n_H}\right)^{1/3},
\end{equation}
where $X=0.76$ is the primordial hydrogen mass fraction, $m_p$ the mass of a proton, and $n_H\approx0.1\ \ccc$ the hydrogen number density in the HII region \citep{Kitayama:05}. Simple conservation of energy arguments allow us to estimate the time to the end of the FE phase,
\begin{equation}
t_{\rm fe}=r_{\rm fe}\left(\frac{M_{\rm ej}}{2E_{\rm tot, POPIII}}\right)^{1/2}.
\end{equation}
Here, $E_{\rm tot,POPIII}$ is the total supernova energy produced by the Pop III stellar population,
\begin{equation}
E_{\rm tot,PopIII} = \sum_{i=1}^{4}N_i E_i,
\label{eq:energy}
\end{equation}
where $N_i$ is the total number of SNe associated with each mass range, as defined in Section~\ref{sec:imf}, and $E_i$ the respective explosion energy (see Table~\ref{tbl:M/E}).

Upon completion of the FE phase, the SNR enters the Sedov-Taylor (ST) phase in which it can be modeled as a point explosion \citep{Sedov:59,Taylor:50}. During this energy-conserving phase, we utilize the well-known scaling for the radius as a function of time, density and energy,
\begin{equation}
r_{\rm st}=\beta\ E_{\rm tot, POPIII}^{1/5}\ \rho_{\rm ISM}^{-1/5}\ t^{2/5},
\label{eq:rstend}
\end{equation}
 where $\beta \simeq1.15$ \citep{Draine:11}, and we take $n_{\rm ISM}\approx 1.0\ {\rm cm^{-3}}$. 
 
\begin{table}
\caption{Energy and fraction of the total stellar mass which is injected into the surrounding medium for each post main-sequence evolutionary outcome adopted, in our model \citep{Karlsson:13}. }
\begin{center}
\begin{tabular}{ccc}
\hline
Type & Energy & Metal yield \\ 
	& [erg]        & ($Y_Z$)	    \\
\hline
\vspace{.1mm}\\
Pair-instability SN & $10^{52}$ & $0.50$ \\
Black hole & $0.0$ & $0.0$ \\
Core collapse SN & $10^{51}$ & $0.10$ \\
Low mass stars & $0.0$ & $0.0$ \\
\hline
\end{tabular}
\end{center}
\label{tbl:M/E}
\end{table}

The ST phase ends when radiative loses become significant, such that energy conservation is no longer a valid assumption.  Based on high resolution simulations with conditions similar to our physical assumptions of the first SN explosions, this occurs at $t_{\rm ST,end}\sim10^5$ yr \citep{Greif:07}. It is at this point that the SNR enters the pressure-driven snowplow phase (PDS), where it reaches into the IGM. At this stage, the evolution is driven by the adiabatic expansion of the hot gas interior to the shell, and will continue until pressure equilibrium is reached between the interior value, $P_i$, and the pressure in the surrounding medium. In this case, we take this medium to be the the IGM, with corresponding pressure $P_{\rm IGM}$.  

To derived the PDS scaling, we start with Newton's second law of motion,
\begin{equation}
\frac{d(m_{\rm sw}v_{\rm sh})}{dt}=4\pi r_{\rm sh}^2 P_i,
\label{eq:force}
\end{equation}
where $m_{sw}$ is the mass ($\propto r_{\rm sh}^{3}$) and $v_{sh}$ is the velocity of the shell as it travels through the ISM.  Given that adiabatic expansion is governed by $PV^\gamma={\rm const.}$, where $\gamma=5/3$, we find that $P_i\propto r_{\rm sh}^{-5}$. For the density scaling, we assume an isothermal profile with $\rho\propto r^{-2}$.  Inserting these expressions into Equation.~\ref{eq:force}, we arrive at the following scaling,
\begin{equation}
r_{\rm pds}=r_{\rm st,end}\left(\frac{t}{t_{\rm st,end}}\right)^{2/5}.
\end{equation}

The SNR will continue to expand until it slows to a speed which is comparable to that of the local sound speed, $c_s=\sqrt{k_{\rm B} T_{\rm IGM}/m_{\rm H}}$. We can estimate the velocity during the PDS phase, 
\begin{equation}
\frac{dr_{\rm sh}}{dt}=v_{\rm sh}\approx\frac{2}{5}\frac{r_{\rm st,end}}{t_{\rm st,end}}\left(\frac{t}{t_{\rm st,end}}\right)^{-3/5}.
\label{eq:drdt}
\end{equation}
Setting $v_{\rm sh}=c_s$,we rearrange Equation~\ref{eq:drdt} to estimate the time at which the PDS phase ends, $t_{\rm final}$, and the corresponding terminal radius of the SNR, $r_{\rm final}$. Here, we assume that $c_s\approx 10\ {\rm km\,s^{-1}}$, which represents the hot phase of the IGM (i.e. $T_{\rm IGM,hot}\approx 10^4\ {\rm K}$), consistent with an IGM heated by an ionization front preceding the shock front (R-type). 

One important omission in the above derivation is the influence of gravity which will act to slow the expanding shell. To account for this, we carry out a simple comparison of the initial explosion energy
to the binding energy of a gas cloud in a typical dark matter minihalo with mass $M_h\approx 10^6\ \Msun$,
\begin{equation}
E_{\rm bind}\simeq\frac{GM_{\rm vir}M_b}{R_{\rm vir}}\,,
\end{equation}
where $M_{\rm vir}=10^6\ \Msun$, $R_{\rm vir}\approx100$ pc (at $z$=15) and $M_b\approx \Omega_b/\Omega_m \times M_{\rm vir}$.  If $E_{\rm tot,PopIII}<E_{\rm bind}$, we take $r_{\rm sh}$ to be equal to $R_{\rm vir}$.  However, if $E_{\rm tot,PopIII}\geq E_{\rm bind}$, we fix the final radius to be at the pressure equilibrium point calculated from Equation~\ref{eq:drdt}.  

\begin{figure}
\begin{center}
\includegraphics[scale=0.32] {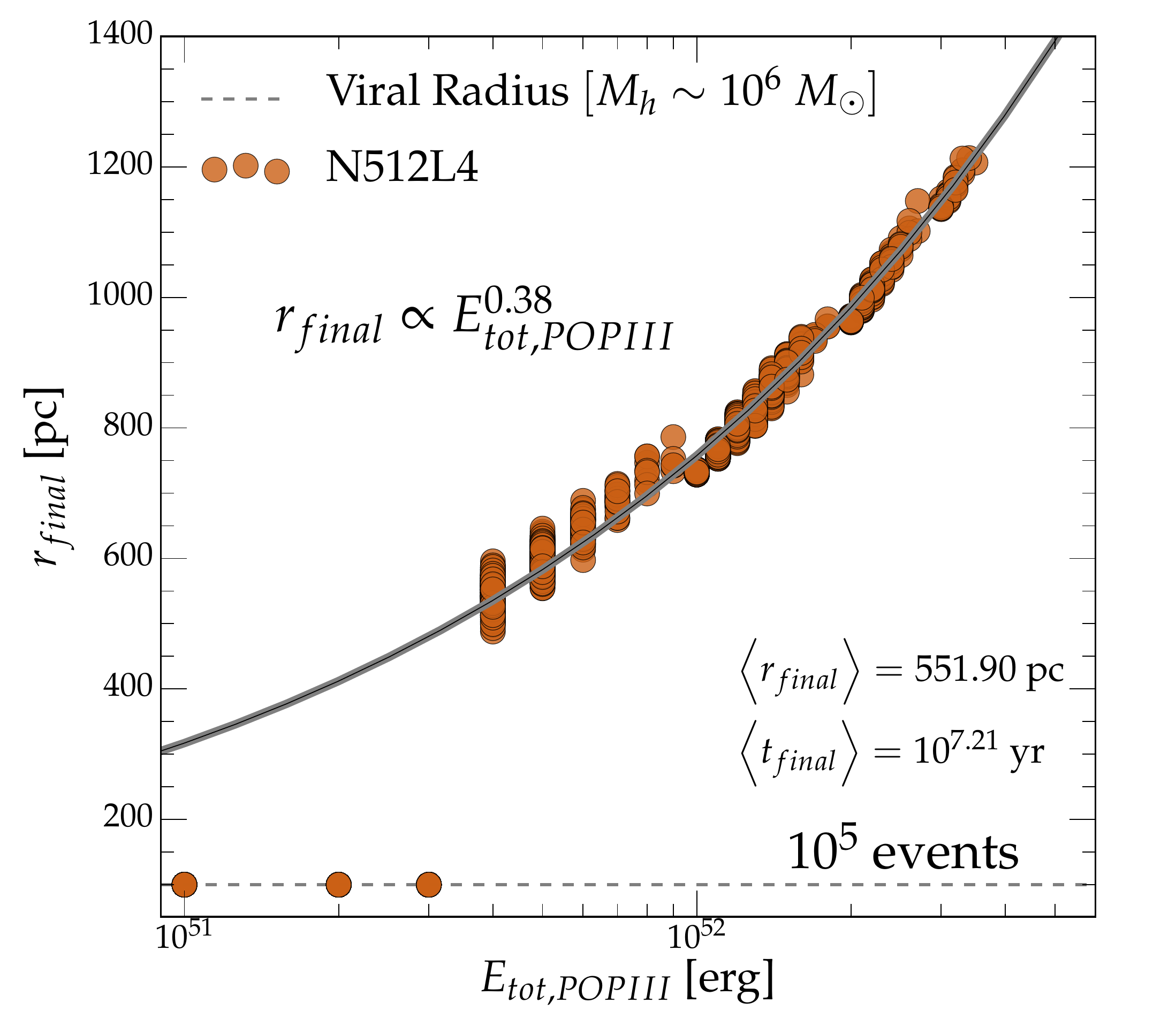}
\caption{Relation between total energy ($E_{\rm tot,PopIII}$) and final enrichment radius ($r_{\rm final}$), derived from a least-square fit (solid line) to the $10^5$ sample (orange circles) stellar populations drawn from the M-Larson IMF.  This relation is utilized in our simulation runs to determine the radius for each unique population.}
\label{fig:erad}
\end{center}
\end{figure}

In Figure~\ref{fig:erad}, we consider the sample of $10^5$ randomly drawn stellar populations from Section~\ref{sec:imf} to derive a relationship between the final radius and total energy injected.  A least-squares fit results in 
\begin{equation}
r_{\rm final}=R_Z\propto E^{0.38}_{\rm tot,PopIII}\,,
\label{eq:final}
\end{equation} 
indicated by the solid line.  This relation is clearly an approximation, given that it has been derived under the assumption of spherical symmetry in the explosion, as well as isotropy and homogeneity in the surrounding medium. Conditions in the actual ISM and IGM of these systems will undoubtedly be far more complicated, widely varying from region to region.  However, we are encouraged by the fact that our mean final radius, $\left<r_{\rm final}\right>\approx 550$ pc, is quite consistent with high resolution, single-event simulations by \citet{MJ:14}.  Therefore, we adopt Equation~\ref{eq:final} as our default model in our simulations for the final radius of enrichment.

\subsubsection{Metallicity Feedback}
Enrichment of metals into the surrounding ISM by a single SF event is determined by summing the contribution from each individual member of the randomly drawn Pop III stellar population:
\begin{equation}
M_{\rm Z,PopIII} = \sum_{i=1}^{4}y_i M_{*,i} ,
\end{equation}
where $M_{\rm Z,PopIII}$ is the metal mass returned to the ISM, $y_i$ is the yield,  and $M_{*,i}$ is the total mass in each mass range, defined in Section~\ref{sec:imf} and Table~\ref{tbl:M/E}.  When a SF event occurs, $Z_{\rm tot,Pop~III}$ is instantly distributed to the surrounding gas contained within a radius of $r_{\rm final}$.  The enriched material is subsequently advected with the general gas motion, thus setting the foundation for the next generation of star formation. 

\subsubsection{Thermal and Ionization Feedback}
\label{sec:ionization}
While the radius of metal enrichment, $R_Z$, is dictated by the physics of the shock front expanding through the ISM, the thermal and ionization impact of the star forming events is governed by the propagation of photons through the ISM/IGM. The corresponding radius can be estimated with an R-type ionization front, which precedes the shock front, via
\begin{equation}
R_{\rm ion}= \left(\frac{3 \dot{N }_{\rm ion}\left<N_*\right>}{4\pi n_{\rm H_I}^2 \alpha_B}\right)^{1/3} \approx 2\ \rm{kpc}.
\end{equation}
This estimation assumes an ionizing flux of $\dot{N }_{\rm ion}\approx 10^{50}\ {\rm s^{-1}}$\citep{Schaerer:02}, an IGM neutral hydrogen density of $n_{\rm H_I}\approx 0.10\ {\rm cm^{-3}}$, and case-B recombination with $\alpha_B=2.59\times 10^{-13}\ {\rm cm^3s^{-1}}$. 

It should be noted that we do not impart kinetic energy to the particles in the form of a velocity/momentum change.  Rather, this energy is injected into the internal energy of each particle (i.e. thermal energy) at the time of star formation, such that
\begin{equation}
T_{i}\approx T_{\rm IGM,Hot}\approx 10^4\ {\rm K}\, .
\end{equation}
This approximation for the interior temperature of the legacy bubble is adopted to be consistent with our radius stalling criterion, i.e. $P_{i}=P_{\rm IGM}$. We also self-consistently assign the degree of ionization by allowing our chemistry solver to update the mean ionization fraction, $\left<x_e\right>$, within $R_{\rm ion}$, based on the thermal energy increase.
Our choice of values is intended to represent the environment left behind after a SN event has expanded and faded away.  Cooling and recombination processes will quickly erase the legacy, imprinted by these parameters \citep{Ritter:12}.  

A consequence of our Legacy model is that a stellar population can form, exclusively containing stars that end their lives as black holes or a combination of black holes and low mass stars ($M_*< 8\ \Msun$).  Both possibilities will leave little to no metallicity imprint on the surrounding environment. They will however emit a large flux of ionizing photons which is deposited within $R_{\rm ion}$. Table~\ref{tbl:rad} summarizes both the metallicity radius and the thermal/ionization radius for each evolutionary outcome.

Our instantaneous enrichment scheme is adopted due to the fact that cosmological volume simulations lack the sub-parsec spatial resolution required to model the expanding shell of a SN remnant through the ISM. This approximation is justified by roughly reproducing the results from select high resolution (sub-parsec) simulations \citep{MJ:14}. 

\begin{table}
\caption{Summary of radii used in our Legacy model for both the chemical and thermal/ionization feedback for each evolutionary outcome.
}
\begin{center}
\begin{tabular}{ccc}
\hline
Type & $R_Z$ & $R_{ion}$ \\ 
	& [pc]        & [kpc]	    \\
\hline
\vspace{.1mm}\\
Pair-instability SN & $\propto E^{0.38}_{\rm tot,POPIII}$ 	& $2.0$ \\
\vspace{.01mm}\\
Black hole 			& $0.0$ 							& $2.0$ \\
\vspace{.01mm}\\
Core-collapse SN 	& $\propto E^{0.38}_{\rm tot,POPIII}$ 	& $2.0$ \\
\vspace{.01mm}\\
Low-mass stars		& $0.0$ 							& $2.0$ \\
\hline
\end{tabular}
\label{tbl:rad}
\end{center}
\end{table}

\subsection{Terminating Pop~III Star Formation}
\subsubsection{Lyman-Werner Background}
\label{subsub:jlw}

The primary coolant ($\Htwo$) which leads to the collapse and subsequent formation of Pop~III stars is highly susceptible to the radiation produced by Pop~III stars.  In particular, molecular hydrogen can easily be dissociated by Lyman-Werner (LW) photons, which have energies in the range $11.2$ eV to $13.6$ eV.  To account for the destruction of $\Htwo$, we implement a self-consistent prescription for a spatially homogeneous LW background with an intensity estimated by,
\begin{equation}
J_{\rm LW,21}(z) \approx 2\left(\frac{\eta_{\rm LW}}{10^4}\right) \left(\frac{\dot{\rho}_*(z)}{10^{-2}\Msun{\rm \ yr^{-1}\ Mpc^{-3}}}\right)\left(\frac{1+z}{10}\right)^3.
\label{eq:jlw}
\end{equation}
Here $J_{21}\equiv J/10^{-21}\ {\rm erg\ s^{-1}\ cm^{-2}\ Hz^{-1}\ sr^{-1}}$, $\eta_{\rm LW}$ represents the number of LW photons produced per stellar baryon and $\dot{\rho}_*$ is the total star formation rate density (SFRD).  To arrive at this normalized equation \citep[e.g.][]{CSS:12}, a stellar population is assumed to last for $\sim5$~Myr, and the IMF to be top-heavy, for consistency with our Pop~III IMF, i.e. $\eta_{\rm LW}\approx1\times 10^4$ \citep{Greif:06}. 

Rates for both photo-dissociation ($k_{\rm di}$) of $\Htwo$ and photo-detachment ($k_{\rm de}$) of ${\rm H^-,\Htwo^+}$ can then easily be calculated by,
\begin{align}
k_{\rm di} &= 1.38 \times 10^{-12} \ \beta\ J_{\rm LW,21} f_{\rm shield,\Htwo}\label{eq:kdi}\\
k_{\rm de} &= 1.10 \times 10^{-10}\ \alpha\ J_{\rm LW,21}\label{eq:kde}\, ,
\end{align}
where $\alpha, \beta$ are parameters that reflect the detailed spectral shape of the
incident radiation (see below). We add these photo-processes to our overall chemical reaction network \citep{Abel:97,Yoshida:03}.  The dimensionless factor $f_{\rm shield,\Htwo}$ accounts for the ability of molecular hydrogen to self-shield against LW photons at high column densities.  At $f_{\rm shield,\Htwo}$=1, the $\Htwo$ is completely unshielded and susceptible to photo-dissociation, whereas at $f_{\rm shield,\Htwo}$=0 it is fully shielded from the LW background.  This self-shielding factor is calculated in our code using the fit provided by \citet{Draine:96}:
\begin{align}
f_{\rm shield,\Htwo} &= \frac{0.965}{(1+x/b_5)^a}\nonumber \\
				 &+ \frac{0.035}{(1+x)^{0.5}}\exp\left[-8.5\times 10^{-4}(1+x)^{0.5}\right],
\end{align}
where $x=N_{\Htwo}/5\times 10^{14}\ {\rm cm^{-2}}$, $b_5=b/10^5\ {\rm cm\ s^{-1}}$, $a=1.1$, and $b=9.12\ {\rm km\ s^{-1}} (T/10^4\ {\rm K})^{1/2}$, the latter representing the velocity spread for the thermal motion of $\Htwo$ \citep{Ahn:07,Wolcott:11}.  The $\Htwo$ column density is estimated on-the-fly in our simulations via $N_{\Htwo}\approx n_{\Htwo}L_{\rm char}$, where $n_{\Htwo}$ is the molecular hydrogen number density and $L_{\rm char}$ is a local characteristic scale length.  To reduce computational overhead, we take $L_{\rm char}\approx L_{\rm J}$ \citep{Bromm:03}, employing the local Jeans length $L_{\rm J}=\sqrt{15k_B T/4\pi \rho G m_H}$. This approach compares favorably to more computationally expensive estimates, such as 'six-ray' or Sobolev approximations. See the discussions in \citet{CSS:12} and \citet{Wolcott:11} for detailed comparisons between methods.

The shape of the assumed spectrum plays an important role in determining the photo-detachment/dissociation rate for ${\rm H^-,\Htwo^+}$.  The values $\alpha=1.71$ and $\beta=0.97$ in $k_{\rm de}$ and $k_{\rm di}$ (Equation~\ref{eq:kdi},\ref{eq:kde}) for Pop III SF are taken from \citet{Agarwal:15}, who use stellar population models \citep[{\scriptsize STARBURST}99,][]{Leitherer:99,Schaerer:02} to generate realistic stellar spectra in order to determine the normalization, rather than blackbody curves at $10^4-10^5$K.  \citet{Agarwal:15} conclude that the $\alpha$ and $\beta$ parameters for Pop~II SF vary depending on the star formation history, metallicity and age of the stellar population.  For this work, we adopt the same $\alpha$ and $\beta$ for Pop~II as Pop~III for simplicity, closely representing a Pop~II starburst with an age of $\sim 20$ Myr and $Z=0.05\ \Zsun$.  In future work, we will consider explicit Pop II star formation, at which time we will be able to determine these values in a more self-consistent manner.

\subsubsection{Pop~II Contribution to LW Flux}
\label{subsubsec:P2P}
Since we are not explicitly forming Pop II stars in our simulation, we must approximate their contribution to the total SFRD and ultimately the global LW flux.  This is accomplished by combining observed values at $z=6-10$ with theoretical estimates for $z>10$ to obtain an approximate functional form for the global Pop~II SFRD.  We find that a simple exponential function of the form
\begin{equation}
\dot{\rho}_{\rm *,PopII} = A e^{-B(1+z)^2},
\label{eq:popII}
\end{equation}
with $A=0.15$ and $B=0.024$, provides a good fit to both observations and theory at $6\leq z \leq 25$.  

In Section~\ref{subsubsec:sfrd}, in conjunction with our SFRD results, we present estimates of the high-$z$ SFRD, derived from the (dust-corrected) UV luminosity function (UVLF), obtained by integrating the UVLF down to a given limit \citep{Finkelstein:16}. Specifically, observations currently reach down to a limit of $M_{\rm UV}=-17$ (green circles). Integrating of the UVLF down to $M_{\rm UV}=-13$ represents an effort to account for systems which are below the observations threshold (gray diamonds). We also consider a theoretical estimate (dashed cyan line) from hydrodynamical simulations of the Pop~II SFRD out to $z\sim25$, found in \citet{Maio:10}. It is evident that our ad-hoc estimation above agrees well with both observations and theory (see Section~\ref{subsubsec:sfrd} for further discussion). It should be noted that the Pop~II contribution is calculated using Equation~\ref{eq:jlw} with $\eta_{\rm LW}\approx 2\times 10^3$ to reflect the Salpeter IMF of Pop~II stars.

\subsubsection{UV Background}
\begin{figure}{}
\begin{center}
\includegraphics[scale=0.40] {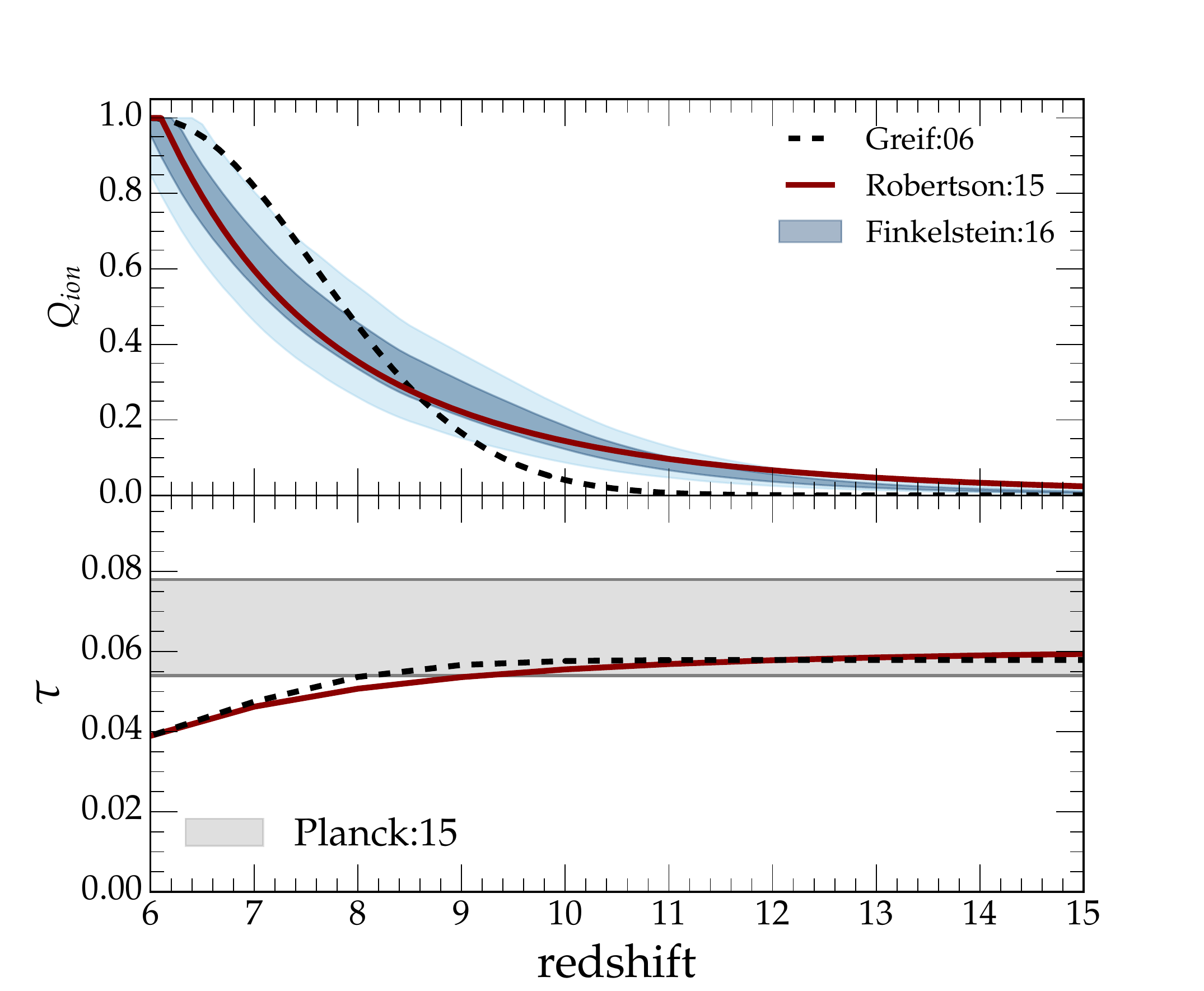}
\caption{{\it Top:} Ionization volume fraction $Q_{\rm ion}$, based on observational constraints \citep{Finkelstein:16}, indicated by the dark/light blue shaded region ($68\%/95\%$ confidence), and \citet{Robertson:15} shown as dark red solid line.  Also shown is the model employed in this work \citep{Greif:06}, where the ionization fraction is represented by a Gaussian at $z\geq 6$. {\it Bottom: } Thomson optical depth, $\tau$, calculated for each of the estimates presented in the top panel, along with the Planck 2015 constraints (gray shaded region).  Here we demonstrate that none of the ionization histories discussed can be ruled out.}
\label{fig:qion}
\end{center}
\end{figure}

The final component to terminating Pop III star formation is the UV ionizing background (UVB) produced by Pop II stars.  This background radiation ionizes gas in both the host and adjacent haloes, effectively suppressing star formation in metal free systems by limiting their ability to effectively cool. Due to the short mean free path of ionizing radiation, the UVB cannot be considered homogeneous, as was the case with the LW background. Since we are not explicitly forming Pop II stars, we adopt a statistical approach to model the patchy nature of the UVB, similar to that employed in \citet{Greif:06}. In our Pop II proxy model (P2P), the probability that a star forming region lies in a region which has been previously ionized can be calculated by
\begin{equation}
p_{\rm ion}=Q_{\rm ion}[1+\xi_{\rm hh}(z)]\,,
\end{equation}
where $Q_{\rm ion}$ is the current cosmic ionization fraction, and the halo correlation function, $\xi_{\rm hh}$, can be interpreted as an excess probability based on the cosmology of our specific simulation volume, see \cite{Greif:06} and references therein for the calculation of $\xi_{\rm hh}$. For the halo mass range encountered in our simulations, $10^6-10^{10}\ \Msun$, we find $\xi_{\rm hh}\sim 10^{-2}$, effectively rendering $p_{\rm ion}\approx Q_{\rm ion}$. We model $Q_{\rm ion}$, following \citet{Greif:06}, with a Gaussian of the form
\begin{equation}
Q_{\rm ion} = \exp(-(z-6)^2/w).
\label{eq:qion}
\end{equation}
Here $w=5$ provides a reasonable representation of the observations (note $Q_{\rm ion}=1.0$ at $z\leq6$).  Each time a gas particle qualifies for Pop III star formation, a pseudo random number, $\mathcal{R}_{\rm ion}$, between [0-1] is drawn. If $\mathcal{R}_{\rm ion}\leq p_{\rm ion}(z)$, the region is considered to be ionized and a Pop~III star is not allowed to form.  We do not explicitly include the ability of dense gas to self shield in our UVB treatment. However, this phenomenon is implicit in the estimates of $Q_{\rm ion}$ upon which our model is based.

The top panel of Figure~\ref{fig:qion} shows several observational estimations for $Q_{\rm ion}$ from \citet{Robertson:15} and \citet{Finkelstein:16}, which are in good agreement with each other.  Our assumed ionization history, generated by Equation~\ref{eq:qion}, is not intended to be an exact fit to the observational estimates, but rather an approximation which gives a reasonable representation of the data which indicates a late reionization scenario.  We note that estimates for the total number of ionizing sources and dark matter halo escape fractions are subject to large uncertainties, rendering determinations of the detailed reionization history of the Universe highly uncertain as well. While the true nature of $Q_{\rm ion}$ is thus uncertain, our choice of a late reionization scenario allows us to provide an upper limit for the amount of metals which can be formed via Pop III star formation.


\section{Results}
\label{sec:results}

We again would like to point out to the reader, that we are only considering metal enrichment from Pop~III star formation. Therefore, in the absence of Pop~II chemical feedback, what we present here is an idealized numerical experiment, providing an upper limit to the ability of Pop~III to enrich the early Universe.

\subsection{Global Properties}
\label{subsec:glob}
\subsubsection{Star Formation Rate Density}
\label{subsubsec:sfrd}

\begin{figure}
\begin{center}
\includegraphics[scale=0.40] {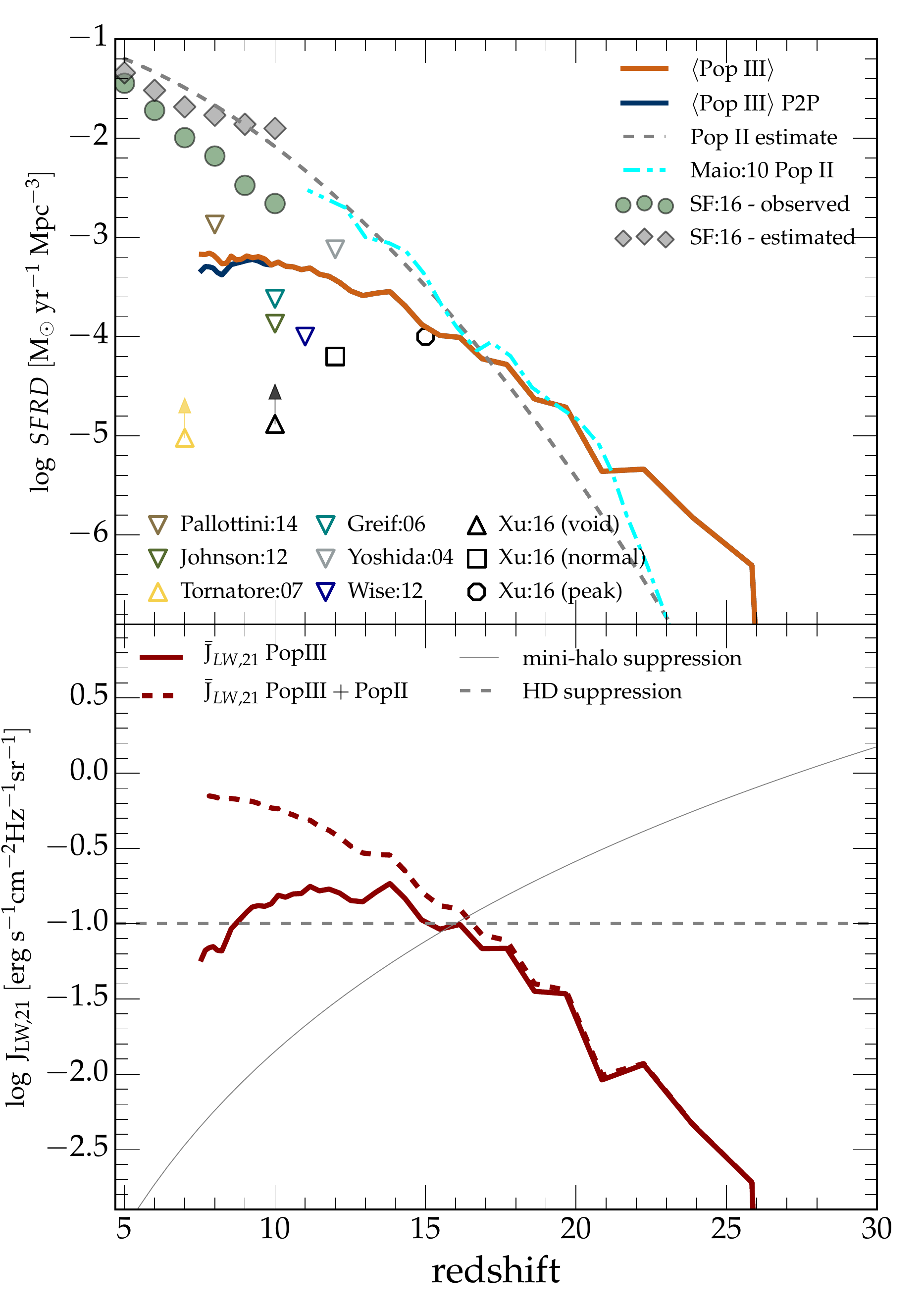}
\caption{{\it Top:} Mean Pop III SFRD from our simulations with P2P on and off (solid blue/orange lines). We also show observations for the Pop II SFRD, derived from the reference UVLF (gray diamonds, green circles) discussed in \citet{Finkelstein:16}, which combines frontier observations $z\geq 5$ from published studies \citep[e.g.][]{Finkelstein:15,Bouwens:15a,Bouwens:15c,McLeod:15,Bowler:14,Schmidt:14,McLure:09,McLure:13,Oesch:13,Oesch:14}, as well as the simulation prediction by \citet{Maio:10} (cyan dashed-dotted line).  The difference between the estimated total (gray diamonds) and observed (green circles) points is that the former are determined by integrating the UVLF down to a limit of $M_{\rm UV}=-13$ to account for systems which are below the current observational threshold.  The gray-dashed line represents our model Pop II SFRD from Equation~\ref{eq:popII}.  We also include comparisons to previous numerical studies from  \citet{Yoshida:04,Greif:06,Tornatore:07, Johnson:13,Wise:12,Pallottini:15,Xu:16b} (gray, teal, yellow, olive, blue, tan and black triangles respectively), which are estimates for the peak of each Pop~III SFRD and the redshift at which it occurs.
{\it Bottom:} Lyman-Werner background flux calculated from Equation~\ref{eq:jlw} for Pop~III only (solid red line) and Pop~III + Pop~II P2P (dashed red line). The solid gray line represents the estimated threshold beyond which $\Htwo$ formation will be suppressed in minihaloes, and the dashed gray line indicates the flux level required to suppress HD.} 
\label{fig:sfrd}
\end{center}
\end{figure}

One of the principal observational constraints that can be placed on Pop~III star formation comes from frontier galaxies observations at $z\gtrsim6$ in the form of the total star formation rate density.  While not directly probing Pop~III star formation, observations of the Pop~II dominated deep-field galaxies \citep[e.g.][]{Finkelstein:15,Bouwens:15a,Bouwens:15c,Bowler:14,Schmidt:14,McLure:09,McLure:13,Oesch:13,Oesch:14} set an upper limit for the combined total of both Pop~II and Pop~III.  In the top panel of Figure~\ref{fig:sfrd}, we present the mean SFRD calculated from our simulation volumes both with (blue) and without (orange) the P2P model enabled, along with estimates of the observed SFRD calculated from a compilation of studies presented in \citet{Finkelstein:16}.  Here we see that both our simulation runs peak and saturate at $\sim10^{-3}\ {\rm M_\odot\ yr^{-1}\ Mpc^{-3}}$ which is roughly an order of magnitude less than the estimated total SFRD at that epoch (gray diamonds).  

We compare our results for the Pop~III SFRD to previous studies which  have employed a variety of numerical techniques, such as AMR \citep{Pallottini:15,Johnson:13}, SPH  \citep{Yoshida:04,Tornatore:07,Johnson:13}, and semi-analytic \citep{Greif:06}.  To make a direct comparison to the results found in different studies we examine the peak of the Pop~III SFRD and the corresponding redshift (see triangles in Figure~\ref{fig:sfrd}).  While the peak redshifts vary, our results broadly agree with those found in \citet{Pallottini:15,Johnson:13,Wise:12,Greif:07,Yoshida:04}.  However, our results are more than an order of magnitude higher than those found in \citet{Xu:16b} and \citet{Tornatore:07}.  In the case of the \citet{Tornatore:07} results, this discrepancy is likely due to the much lower halo mass resolution ($M_h\sim10^8\ \Msun$) utilized in their study which did not allow for Pop~III star forming regions to be fully resolved.  The \citet{Xu:16b} work utilized highly resolved AMR simulations to look at star formation in void, normal, and peak density regions. Our results are only significantly different when compared to the SFRD in the void regions.

Due to the P2P model, the blue line in Figure~\ref{fig:sfrd} begins to deviate to lower values at $z\sim10$.  This behavior is expected as the P2P model is directly coupled to the ionization fraction utilized which begins to ramp up at the same redshift.  From Figure~\ref{fig:qion} and Equation~\ref{eq:qion} it can be seen that $Q_{\rm ion}\approx0.64$ at $z=$7.5.  While resulting in only a small variation at the terminal redshift reached in our simulation, the rapid increase of the ionization fraction leads to a mostly ionized Universe by $z$=6.  The increased ionizing flux is directly responsible for terminating Pop~III star formation as it heats the gas which cannot self-shield in low density, low metallicity haloes.  Due to computational constraints, we were not able to reach $z$=6 with this suite of simulations. However, it is plausible to predict that our P2P-enabled simulation SFRD would continue to decline until being terminated at $z\approx$6.  We have completed lower resolutions runs to confirm that our P2P model does have the predicted impact on the global Pop~III SFRD.

We note that all further analysis will be conducted on the simulation run with P2P enabled.  It is also important when examining these results to recall that we are only considering metal enrichment from Pop~III star formation.

\subsubsection{Metallicity}
\label{subsubsec:metals}

\begin{figure*}
\begin{center}
\includegraphics[scale=0.42] {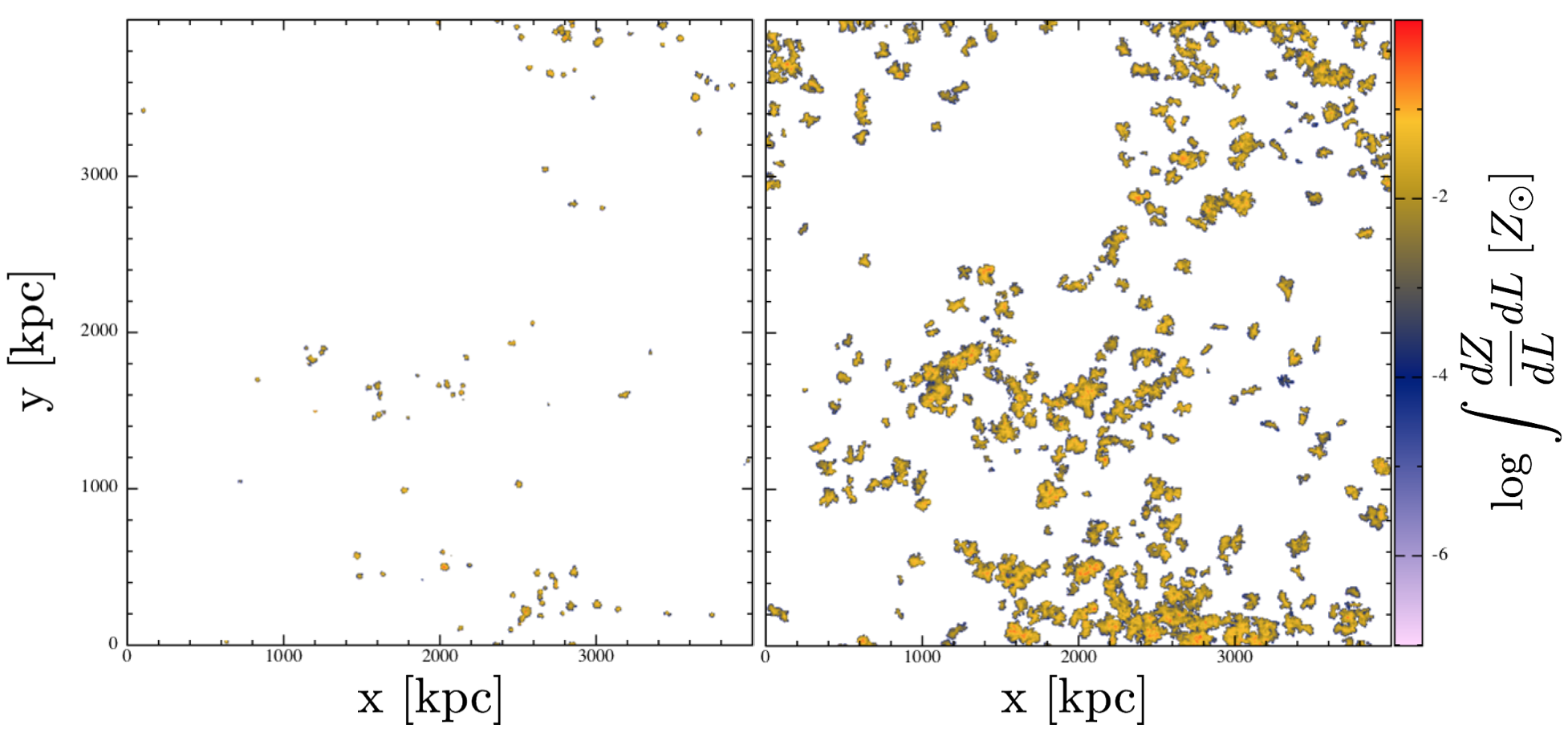}
\caption{  Metallicity projection plot for $z$=15 (left) and $z$=7.5 (right) for our N512L4\_P2P run.  The color bar represents the integrated metallicity along the line of sight through the z-axis of the computational volume.  Visualizations generated using {\small SPLASH} \citep{Price:07}.   
} 
\label{fig:viz}
\end{center}
\end{figure*}

\begin{figure}
\begin{center}
\includegraphics[scale=0.34] {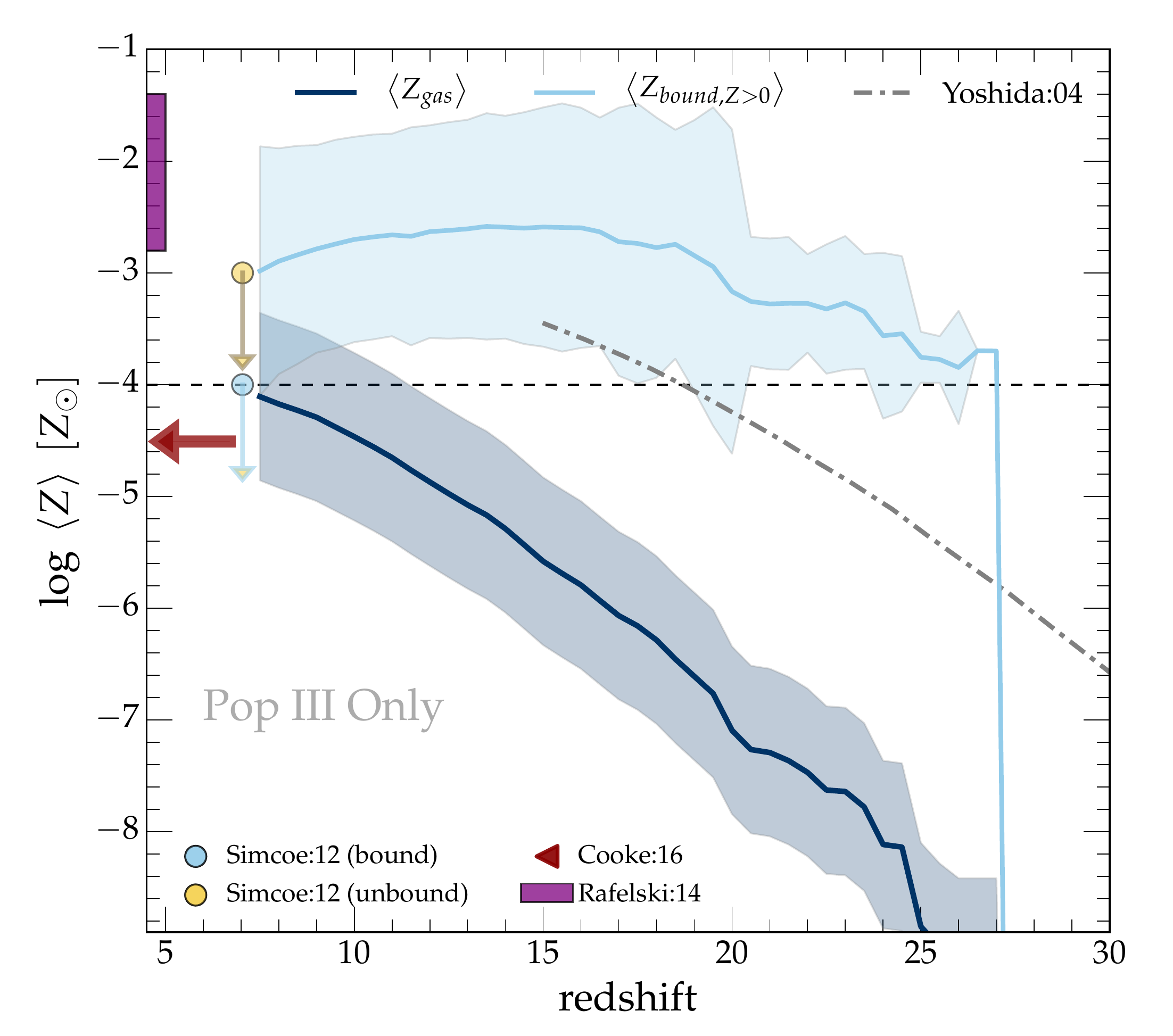}
\caption{Redshift evolution of the mean metallicity for all gas particles in our simulation volume (solid blue line), and for haloes which have $Z>0$ (cyan). The horizontal dashed gray line represents the critical metallicity ($Z_{\rm crit}=10^{-4}\Zsun$) for the Pop~III to Pop~II transition. We also reproduce the predicted metallicity evolution from \citet{Yoshida:04}, where {\it all} Pop~III stars were considered to end their lives as PISNe, thus imparting a maximum amount of metals back into their environment. Observations of high redshift DLA systems are indicated by the yellow/blue circles \citep[$z=7.04$;][]{Simcoe:12}, and the purple range \citep[$z>4.7$;][]{Rafelski:14}. The red arrow points to a lower redshift observation by \citet{Cooke:16}.  The tan circle is the estimate based on the assumption of an unbound medium, whereas the yellow point is for a bound structure.  Both are considered upper limits.  
} 
\label{fig:metal}
\end{center}
\end{figure}

The primary legacy left by Pop~III star formation, and the main focus of this study, are the metals which are imparted to the ISM/IGM by Pop~III SNe. In Figure~\ref{fig:viz} we present a visualization of our simulation volume at $z$=15 (left) and $z$=7.5 (right) to illustrate the patchy nature of Pop~III enrichment.  In Figure~\ref{fig:metal}, the solid blue line represents the evolution of the mean metallicity of all the gas particles in our simulation volume.  As the total amount of baryonic matter in our simulation box is finite, the mean metallicity will continue to rise, as long as Pop~III star formation continues. Furthermore, once Pop~III star formation is terminated the mean metallicity will remain constant. Interestingly, this value is only approaching the critical metallicity required for the Pop~III to Pop~II transition, $Z_{\rm crit}=10^{-4}\ Z_\odot$, by $z\sim 7$ (dashed gray line). 

Simple extrapolation of our metallicity value at $z\geq7.5$ leads to an average metallicity which is only slightly greater than $Z_{\rm crit}$ by $z=6$.  This result is in contrast to the assumptions made by  previous generation large cosmological volume simulations \citep[e.g.][]{Jaacks.etal:12a,Jaacks.etal:12b} in which Pop~III stars are not explicitly formed and a metallicity floor for all gas particles is implemented at $z>15$ (typically the critical value of $Z\approx  10^{-4}\ Z_\odot$). The consequence of this approach is that Pop~II stars could be allowed to form in regions which otherwise would be devoid of metals, thus potentially overproducing Pop~II stars in $z\geq6$ galaxies.  We intend to further explore this conjecture in future work.

The fact that our mean metallicity does not exceed $Z_{\rm crit}$ prior to reionization suggests that Pop~III star formation is not a globally self-terminating process, as previously argued by \citet{Yoshida:04}. The results of the \citet{Yoshida:04} study is indicated by the gray dashed-dotted line in Figure~\ref{fig:metal}, substantially higher than our enrichment history. The large difference is a direct consequence of the Pop~III IMF assumed in that earlier work, which was chosen to be very top-heavy, considering only stars with $M_*=100-300\ \Msun$.  The impact of this is that each SN event was considered to be a PISN, which has a much higher mass yield than other SN types (see Table~\ref{tbl:M/E}).  Therefore, the \citet{Yoshida:04} results can be viewed as an upper limit to the mean metal enrichment. 

It should be noted that, in the context of this work, we consider the concept of global self-termination to refer to the ability of Pop~III star formation to suppress future Pop~III star formation throughout the simulation volume rather than only in a given host halo.  For bound haloes we find that metal enrichment rapidly increases to above $Z_{\rm crit}$ shortly after Pop~III star formation begins, as indicated by the solid cyan line in Figure~\ref{fig:metal}. We find that within bound haloes there is a plateau metallicity value of $Z/\Zsun\approx10^{-3}$, consistent with previous AMR simulations from \citet{Wise:12}.

The currently highest-redshift observation to constrain our modeling comes from \cite{Simcoe:12} in the form of a damped Lyman alpha (DLA) absorption system at $z\simeq7$. This system has an estimated metallicity of $Z=10^{-4}\ \Zsun$, if residing in a bound proto-galactic halo, or $Z=10^{-3}\ \Zsun$, if part of the diffuse, unbound IGM. The mean metallicity from enriched, bound systems found in our simulation volume is nearly one dex higher than the observed estimate (cyan circle). However, the observed point does fall within our 95\% confidence range, indicating that, while very rare, such systems do exist in our computational box.  In Sec~\ref{subsec:igm}, we sample such a region to create a mock observation.  Note that we define ``bound`` to be gas which resides within the virial radius of a dark matter halo which is consistent with \citet{Simcoe:12}.

The \citet{Rafelski:14} observations at $z\simeq 4.7$ (purple range) contain systems with a range of metallicities ($10^{-1.4}\lesssim Z/\Zsun \lesssim 10^{-2.8}$); all of which are higher than our mean value for bound systems.  This is expected as these observations most likely contain enrichment from both Pop~III and Pop~II star formation.  This provides a strong upper bound for Pop~III only enrichment.

\subsubsection{Volume Filling Fraction}
\label{subsubsec:vff}

\begin{figure}
\begin{center}
\includegraphics[scale=0.37] {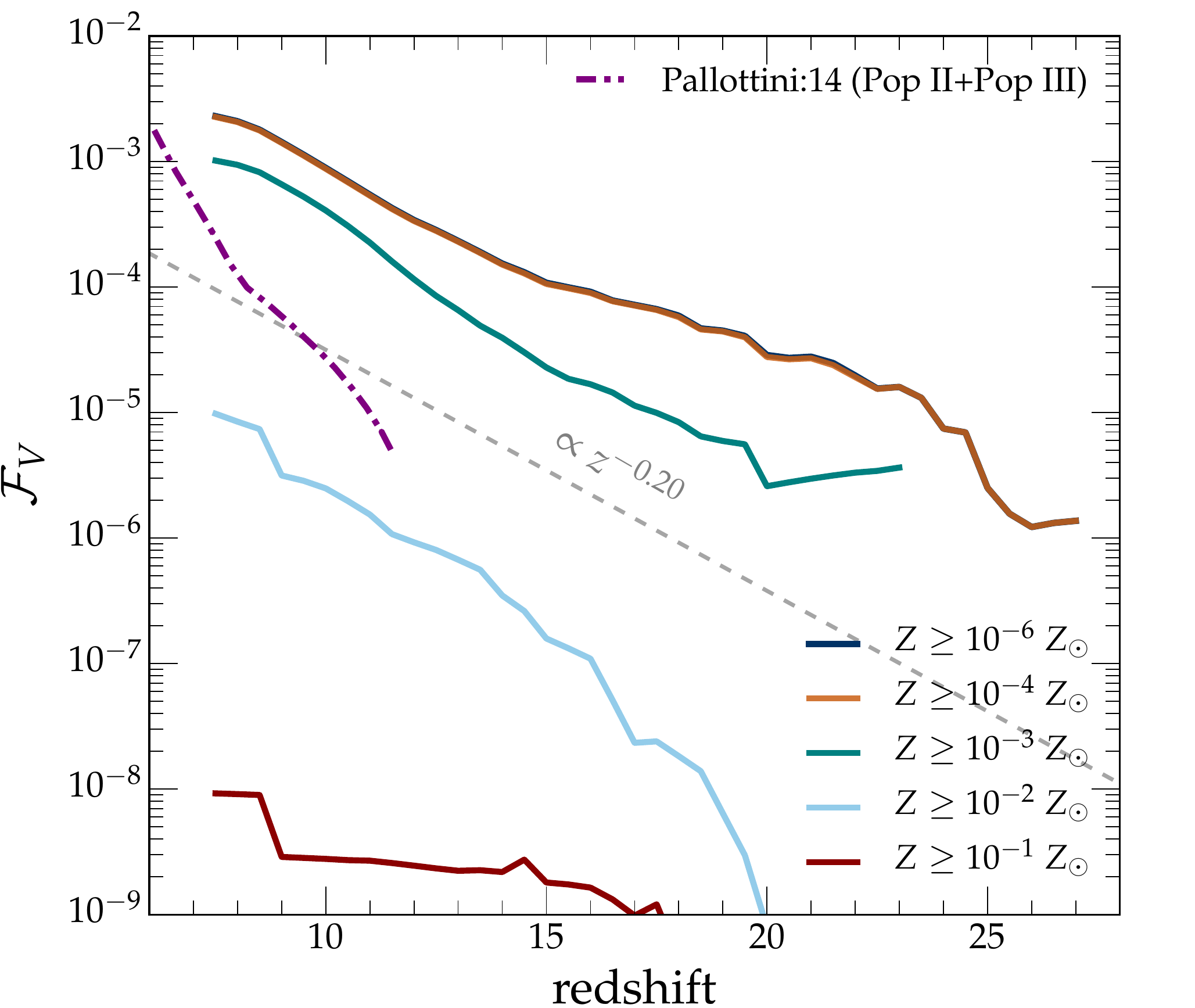}
\caption{Volume filing fraction ($\mathcal{F}_V$) of our simulation box which is enriched to at least the values indicated.  From top to bottom, the lines represent $\log Z/Z_\odot > -6, -4, -3, -2, -1$ (blue, orange, teal, cyan, red).  The dark blue and orange lines essentially overlap with only minor variation, indicating that our enrichment floor of an individual region is $\sim 10^{-4}-10^{-5}\ Z_\odot$. The gray dashed line indicates a growth rate of $\log \mathcal{F}_V\propto z^{-0.20}$. It should be noted that the gray line is not an exact fit to the data.  The purple dash-dotted line represents an estimate for $\log Z/Z_\odot >-2$ from \citet{Pallottini:15}, which includes metals from both Pop~II and Pop~III star formation. }
\label{fig:vfrac}
\end{center}
\end{figure}

In Figure~\ref{fig:vfrac} we explore the redshift evolution of the fraction of our simulation volume which is enriched by Pop~III star formation to at least the indicated values (i.e. volume filling fraction).  At lower redshifts, $z\leq20$, gas which is enriched to $Z>10^{-2}-10^{-3}\ Z_\odot$ can be considered to reside in star forming regions, while gas enriched to $Z<10^{-3}\ Z_\odot$ is associated with the IGM.  While not identical, the blue line representing gas with $Z\geq10^{-6}$ and the orange line representing gas enriched to  $Z\geq10^{-4}$ are nearly indistinguishable over the entire simulation run.  Due to our mass and spatial resolutions we are unable to sufficiently resolve metallicities of $Z<10^{-6}$.  However, our results suggest that even minimal Pop~III star formation activity quickly enriches regions to $Z\geq10^{-6}\ Z_\odot$. In contrast, the appearance of regions which are  enriched to $Z\geq10^{-2}\ Z_\odot$ is delayed by $\sim 70$ Myr from the start of our Pop~III star formation (i.e. $z=27-20$) due to the time required to build up metals in any one given star forming region.  Additionally, an upper limit to the Pop~III metallicity manifests itself as there is a large ($\sim 3$ dex) drop between $Z\geq10^{-2}\ Z_\odot$ and $Z\geq10^{-1}\ Z_\odot$, with no gas enriched to $Z_\odot$.

The fraction of our volume which is enriched to $Z\geq10^{-6}$ shows an exponential growth trend ($\log \mathcal{F}_V\propto z^{-0.20}$) from a value of $\sim10^{-6}$ at $z$=27 to $\sim10^{-2.7}$ at $z$=7.5.  The approximately equivalent growth rate over all metallicities shown in Figure~\ref{fig:vfrac} can be directly attributed to the growth in the Pop~III SFRD which demonstrates a similar growth trend (SFRD$\propto z^{-0.20}$) over the majority of the simulation run.  Any deviations from this behavior in the volume fraction are likely the result of merger events.

Comparison of our Pop~III volume filling fraction to previous numerical studies is difficult as the majority do not differentiate between Pop~II and Pop~III enrichment \citep[e.g.][]{Oppenheimer:06,Tornatore:07,Johnson:13,Pallottini:15} . However, just as with the observed SFRD (see Sec.~\ref{subsubsec:sfrd}), past numerical studies can be utilized to provide upper limits for the metal enrichment. Specifically, the purple dash-dotted line in Figure~\ref{fig:vfrac} represents results from \citet{Pallottini:15}, who utilized  AMR simulations to study early cosmic enrichment.  It is clear that our result, for the same enrichment value of $Z\geq10^{-2}\ Z_\odot$ (solid cyan line), is considerably lower, which is expected as we here include only Pop~III enrichment.

\subsection{Virialized Structures}
\label{subsec:vir}

\begin{figure}
\begin{center}
\includegraphics[scale=0.38] {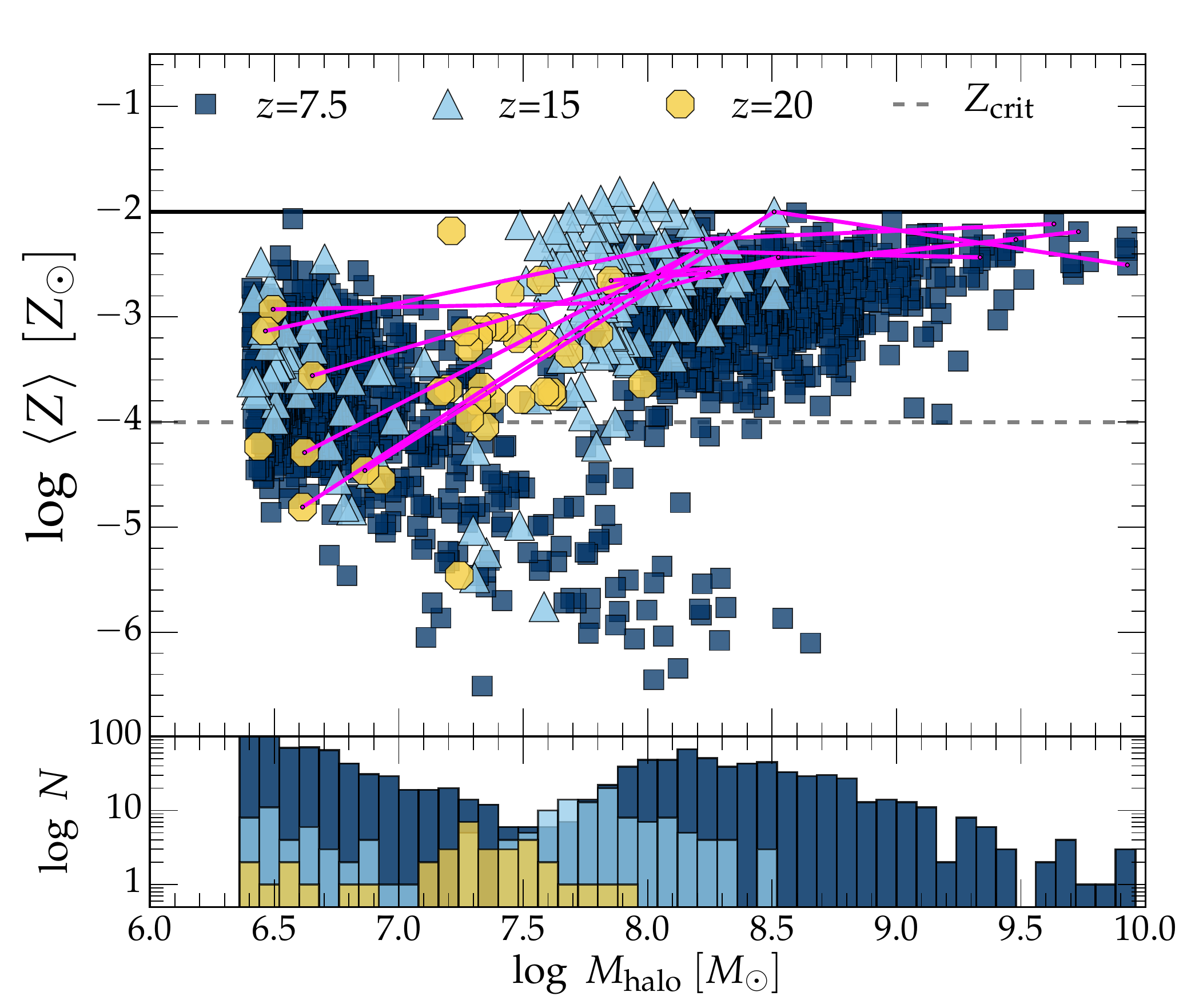}
\caption{{\it Top:} Average halo metallicity as a function of dark mater halo mass for redshifts $z=20, 15, 7.5$ (yellow octagons, cyan triangles, blue squares).  
We also display the evolutionary paths of select individual systems, indicated by the solid magenta lines. {\it Bottom:}  Histogram of the enriched dark matter halo mass distribution for each redshift above.} 
\label{fig:mhZ}
\end{center}
\end{figure}

By grouping gas particles associated with their host dark matter haloes, we can extract information regarding the evolution of metals within bound, virialized structures.  In the top panel of Figure~\ref{fig:mhZ} we present the relation between dark matter halo mass and mean metallicity for three different redshifts ($z$=20, 15, 7.5).  
It is clearly seen that a strong relation between the mass of the halo and it's mean metallicity is established by $z\simeq 7.5$. This trend is expected as higher mass haloes experience higher star formation rates, and therefore more metal enrichment. Specifically, those haloes will continue forming Pop~III stars through accretion of pristine gas, which is processed at a higher efficiency. It has indeed been shown by \citet{Greif:08} that higher mass haloes host Pop III star formation at increased efficiency due to their deeper potential wells, leading to a larger number of free electrons. The latter then catalyze enhanced production of H$_2$, followed by increased HD formation, resulting in more effective cooling. Thus, those haloes exhibit higher metallicities, as shown in Figure~\ref{fig:mhZ}.

In our P3L model, star formation is only allowed for gas with $Z<Z_{\rm crit}$. However, it is clear from Figure~\ref{fig:mhZ} that bound systems are able to enrich well beyond the critical value.  This can be attributed to the constant accretion of primordial metal free gas onto the halo which, once cooled, can fuel new star formation.  In conjunction, haloes also acquire metals through mergers with systems which have previously been enriched. 

An interesting feature found when examining the halo mass to mean-metallicity relation is the heavy element ``Pop~III plateau'' of $\left<Z\right>\sim 10^{-2}\ Z_\odot$, beyond which there are only a few haloes at $z$=15 and none at $z$=7.5.  This is supported by the results found in Section~\ref{subsubsec:vff}, where only a tiny fraction ($<10^{-5}$) of any region in our simulation volume was enriched to $Z\geq10^{-2}\ \Zsun$ by $z$=7.5. These trends suggest that any bound halo with gas estimated to have $\left<Z\right> >10^{-2}\ Z_\odot$ would have to contain metals generated from Pop~II star formation. Our ``plateau'' value of $\left<Z\right>\sim 10^{-2}\ Z_\odot$ is very similar to that found in \citet{Wise:12}, where $\log \left<Z\right> = -2.1\ Z_\odot$, and reflects a natural balance point between metal production and dilution. It is also noteworthy that a metallicity floor emerges as we have zero bounds systems that are enriched to $Z<10^{-7}\ \Zsun$, although this may in part reflect our limited mass resolution (see the discussion in \cite{MJ:17}).

A second interesting feature in Figure~\ref{fig:mhZ} is the ``gap'' at $M_h\approx10^{7.5}\Msun$, which appears in the top panel and in the histogram at the bottom of the same figure, resulting in a bi-modal distribution at $z$=7.5 (dark blue).  It should be noted that there is no dearth of haloes at this mass, as can be seen in the total halo mass distribution (dark solid blue bar below $\log Z/Z_\odot=-7$).  Rather, there are few haloes at this mass which are enriched.  Therefore, it is more appropriate to characterize this deficit as a "metallicity gap".  

The majority of the population with $M_h\lesssim10^{7.5}\Msun$ at $z$=7.5 is enriched by external processes, defined in this work to be any process other than in-situ star formation (i.e. accretion of enriched gas, star formation in nearby haloes or mergers). These external enrichment haloes make up $\sim45\%$ of the entire enriched population at $z$=7.5 (blue squares with red border). At higher redshift, the $M_h\lesssim10^{7.5}\Msun$ haloes are a combination of 'in-situ' and externally enriched systems. Due to hierarchical structure formation and their proximity to star forming haloes, they quickly merge with the more massive systems, which can be clearly discerned in the evolutionary tracks in Figure~\ref{fig:mhZ}, indicated by the magenta lines connecting progenitors to descendants. A similar effect was found in \citet{MJ:17}, who simulated the assembly of fossil dwarf galaxies in the Local Group.

\subsection{Diffuse IGM}
\label{subsec:igm}
\begin{figure}
\begin{center}
\includegraphics[scale=0.37] {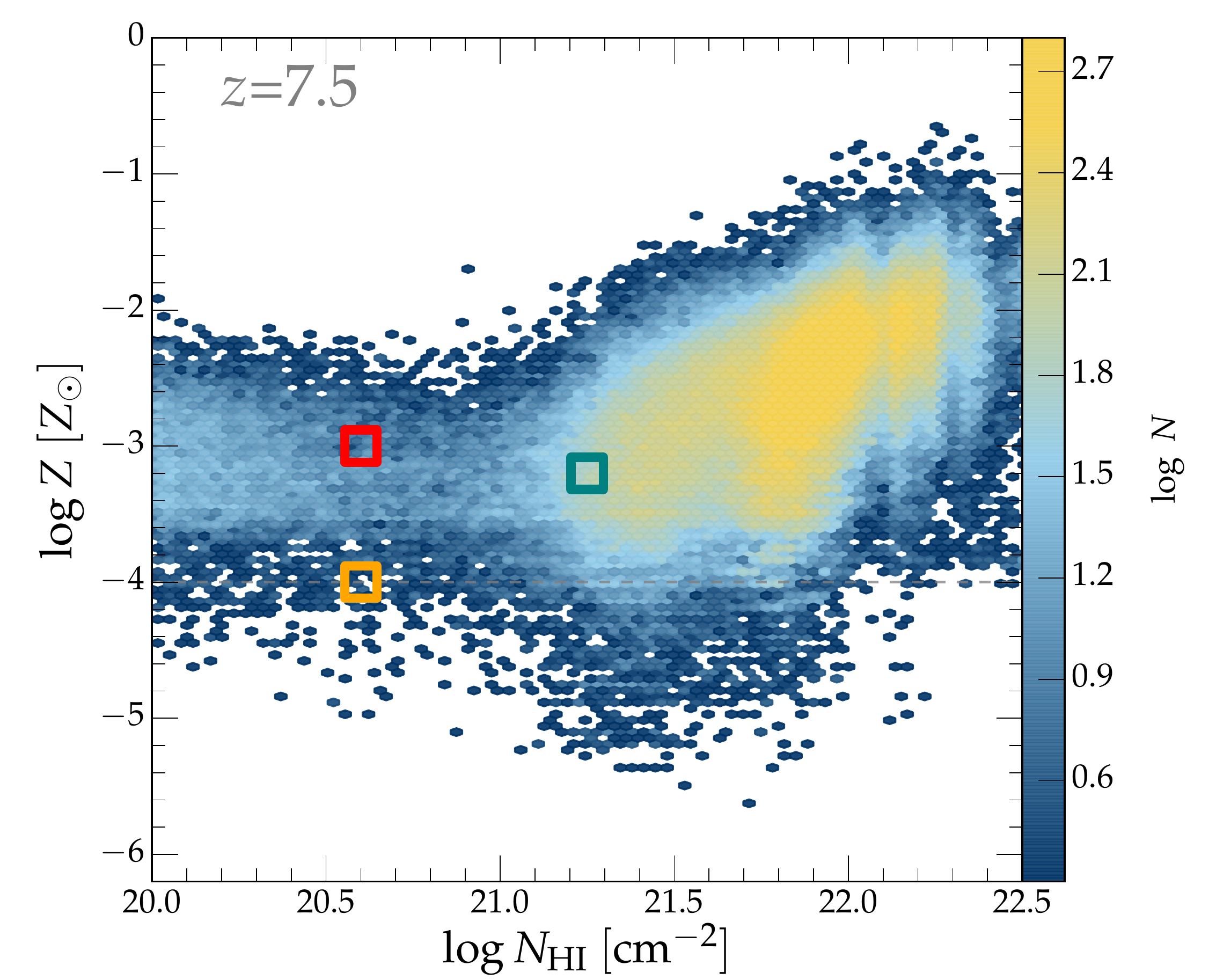}
\caption{Gas particle metallicity as a function of each particles estimated column density at $z=$7.5.  The color scale indicates the number density of the gas particles in this phase space.  It should also be noted that, due to the large particle count, only a representative subset (1/10) of particles are plotted here. We have verified that this sampling does not impact the underlying relation. The orange, red and teal squares correspond to regions which were sampled to create mock absorption-line spectra (see Figure~\ref{fig:spec}).} 
\label{fig:colZ}
\end{center}
\end{figure}

DLA systems are valuable probes of the CGM/IGM temperature, density and metal structure, as photons from bright background sources, such as AGN and GRB afterglows, are absorbed while  propagating through these systems. Given that the temperature, density and metal content are intrinsic properties of the gas in our simulations, it is possible to generate mock observations to probe them. To accomplish this, we consider the radiative transfer equation, simplified for a purely absorptive to  
\begin{equation}
I_\nu = I(0) e^{-\tau_\nu}\, ,
\end{equation}
valid for a static medium. Here, $I_\nu$ is the specific intensity, $I(0)$ the emission from the background source, and $\tau_\nu$ the optical depth along the line of sight (LOS) between the system and observer. For simplicity, we adopt a simple power-law, $f_{\nu}\propto\nu^{-1}$, representing a GRB afterglow as our background source spectrum, such that there are no intrinsic spectral features imprinted by the local environment. To normalize the afterglow emission, we adopt the source parameters employed in \cite{Wang:12}.  Specifically, the GRB is assumed to originate in a minihalo at $z\simeq 16.5$, observed at the reverse shock
crossing time \citep[see figure 10 in][]{Wang:12}.

The usual procedure for obtaining the optical depth from a simulation is to create a column along the LOS, and integrate $d\tau_\nu=\sigma_\nu n_X dl$.  Here $\sigma_\nu$ is the frequency dependent cross-section, $n_X$ is the number density of a given species, and $dl$ is a differential length along the LOS. Establishing an appropriate column is somewhat arbitrary in the context of a cosmological volume simulation. We therefore employ an idealized local prescription, where we simplify our optical depth calculation to 
\begin{equation}
\tau_\nu=\sigma_\nu n_X L_{\rm char}= \sigma_\nu N_X\,.
\label{eq:tau}
\end{equation}
Here, we take $L_{\rm char}$ to be the local Jeans length of the gas particle (see Sec~\ref{subsub:jlw} for additional discussion regarding $L_{\rm char}$), and $N_X$ is the column density of the absorbing species. This prescription allows us to create synthetic absorption spectra by simply sampling the column density and metallicity from Figure~\ref{fig:colZ}.  

\begin{figure}
\begin{center}
\includegraphics[scale=0.37] {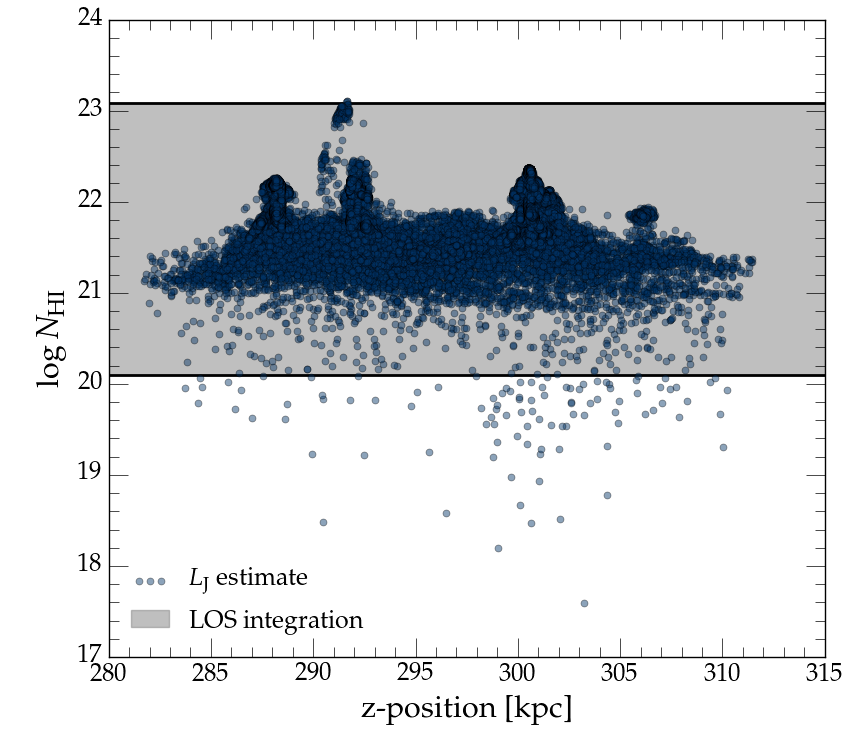}
\caption{ Comparison between column densities calculated for a single halo extracted from our simulation volume using LOS integration (gray horizontal span denotes the minimum and maximum range) and the approximation presented in Section~\ref{subsec:igm}, which utilizes the Jeans length for a given SPH particle (blue circles). The $z$ coordinate marks positions of individual SPH particles along the LOS.
} 
\label{fig:NHIcomp}
\end{center}
\end{figure}

\begin{table}
\begin{center}
\caption{Elements considered in creating the absorption spectra in Figure~\ref{fig:spec}, along with their metal abundance ($X_x=m_x/m_*$), rest-frame wavelength and oscillator strength. Metal yields were take from \citet{Heger:10} for the zero metallicity Type~II case, and \citet{Heger:02} for the PISN case. Atomic transition data was obtained from \citet{Morton:03}.
}
\begin{tabular}{cccccl}
\hline
Element & $X_{x,{\rm TypeII}}$ & $X_{x,{\rm PISNe}}$  & State & 	$\lambda_{\rm rest}$	& $f_{\rm osc}$	\\  
&$10^{-2}$	&$10^{-2}$	& 	&	[\AA] &	\\
\hline
\vspace{.05mm}\\
C	&	$1.08$	 & $4.92$	& \ion{C}{ii}	& $1334.5$ & $0.1278$\\
	&			 & 			& \ion{C}{iv}	& $1548.2$ & $0.1908$\\
	&			 & 		 	& \ion{C}{iv}	& $1550.8$ & $0.0952$\\
O	&	$8.80$	 & $3.51$	& \ion{O}{i}	& $1302.2$ & $0.0489$\\
Mg	&	$0.52$	 & $2.01$	& \ion{Mg}{ii}	& $2796.4$ & $0.6155$\\
Si	&	$0.750$	 & $0.225$	& \ion{Si}{ii}	& $1260.4$ & $1.190$\\
	&			 & 			& \ion{Si}{ii}	& $1304.4$ & $0.094$\\
	&			 &	 	 	& \ion{Si}{iv}	& $1393.8$ & $0.514$\\
    &			 &	 	 	& \ion{Si}{iv}	& $1402.8$ & $0.2553$\\
Fe	&	$1.10$	 & $3.08$	& \ion{Fe}{ii}	& $1608.5$ & $0.058$\\
	&			 & 			& \ion{Fe}{ii}	& $2344.2$ & $0.114$\\
	&			 &	 		& \ion{Fe}{ii}	& $2382.8$ & $0.300$\\
    &			 &	 		& \ion{Fe}{ii}	& $2586.7$ & $0.069$\\
    &			 &	 		& \ion{Fe}{ii}	& $2600.2$ & $0.239$\\
\hline
\end{tabular}
\label{tbl:lines}
\end{center}
\end{table}

Evidently, our approach is only valid if the cosmological redshift across our simulation box is negligible, $\Delta \lambda/\lambda_0 = L_{\rm phys} H(z)/c \ll 1$. At $z$=7.5, we find $\Delta \lambda/\lambda_0 \approx 1.5\times 10^{-3}$.  Therefore, even if two systems along a line of sight were separated by the entire box length, the $\Delta\lambda$ is so small that existing high-resolution spectrographs would be unable to resolve the line shift. Further, our method for estimating column density, using locally evaluated quantities, is clearly very approximate, and therefore should not be considered a replacement for precise LOS determinations. In Figure~\ref{fig:NHIcomp}, we present the results of a test in which we extract a single halo from our simulation volume and compare the column densities calculated using direct LOS integration, intersecting the system at different impact parameters, with the local, particle-by-particle estimates adopted in this work. This test confirms that our idealized method provides reasonable agreement over the range of densities studied here.

The final ingredient required to calculate the optical depth is the frequency dependent cross-section,
\begin{equation}
\sigma_\nu = \sqrt{\upi}e^2/m_ec f_{\rm osc} H(u,x)/\Delta \nu_D\, ,
\label{eq:sigma}
\end{equation}
where $f_{\rm osc}$ is the oscillator strength, $H(u,x)$ the Voigt function, and $\Delta\nu_D$ the Doppler width. We refer the reader to \citet{Wang:12}, and references therein, for details of the radiative transfer calculation, and the parameter choices in Equation~\ref{eq:sigma}. Data for the atomic transitions are taken from \citet{Morton:03}.

\begin{figure}
\begin{center}
\includegraphics[scale=0.20] {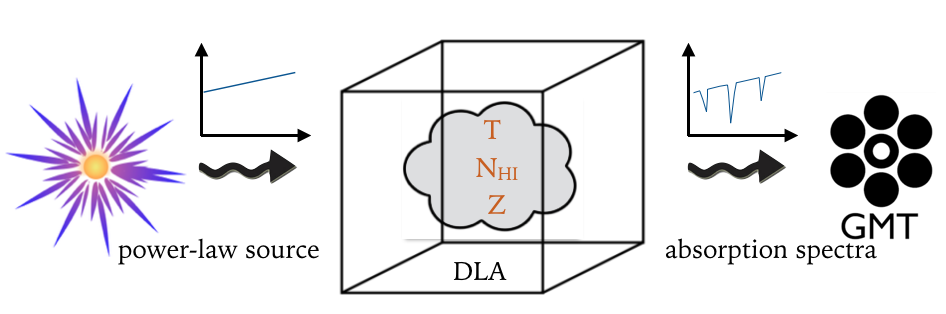}
\includegraphics[scale=0.30] {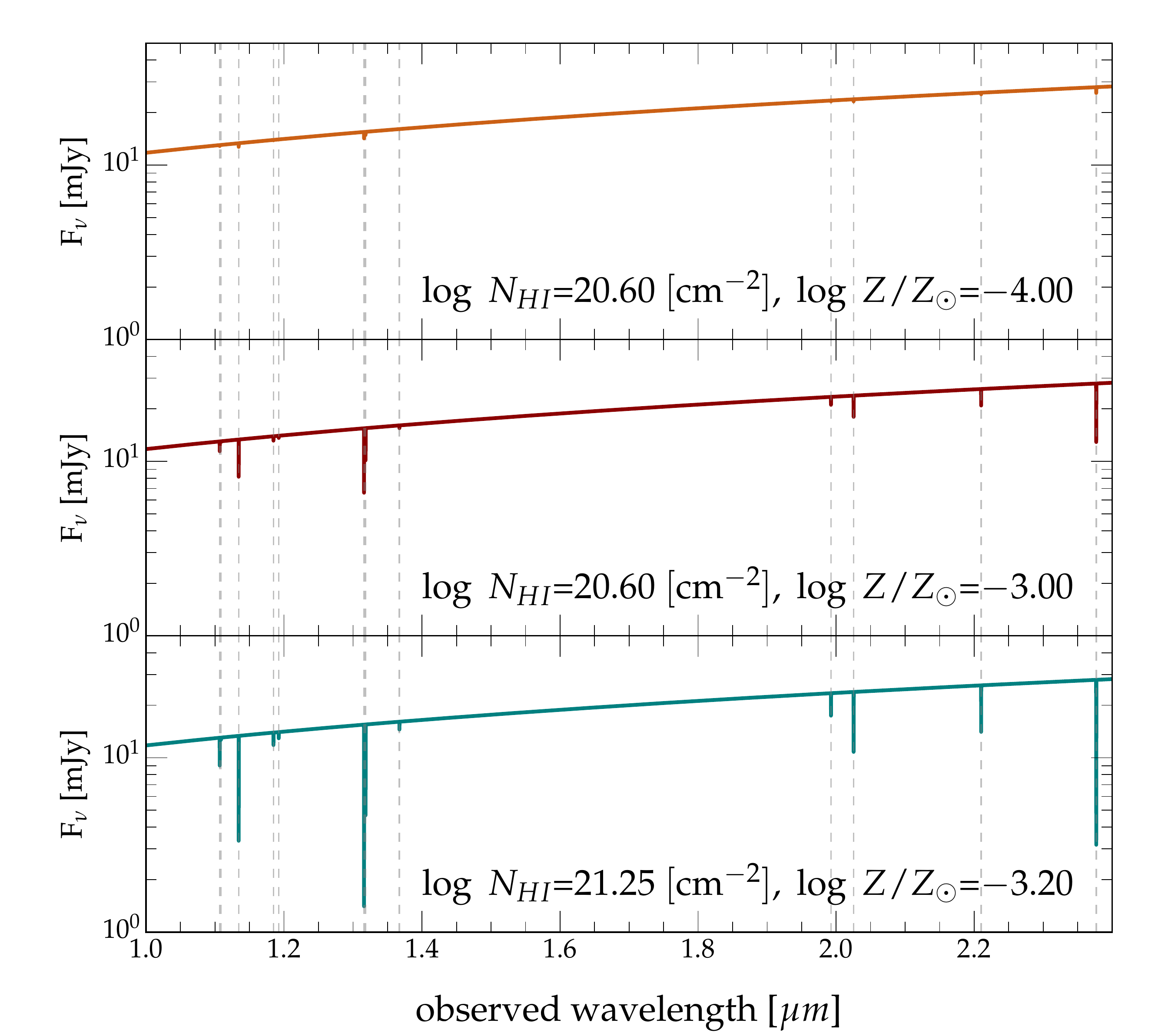}
\caption{{\it Top:} Illustration of our scenario with a generic GRB power-law spectrum passing through an IGM absorption system with a unique set of physical properties ($T, N_{\rm HI}, Z$), before reaching the observer. {\it Bottom:} A sample of three synthetic spectra, generated from our simulation volume. The top two are representative of the \citet{Simcoe:12} estimates, whereas the bottom is a randomly selected region from the parameter space in Figure~\ref{fig:colZ}. Temperatures of the three regions are very similar with $T\approx10^{3.8}$ K, corresponding to a Doppler width of $b\approx10\ {\rm km\,s^{-1}}$.} 
\label{fig:spec}
\end{center}
\end{figure}

\begin{table}
\begin{center}
\caption{Observed frame equivalent widths ($W$) for select ions calculated from the spectra presented in Figure~\ref{fig:spec}.  The subscripts $1, 2, 3$ correspond to the spectra top to bottom.  The final column $W_{\rm obs}$ contains the values presented in \citet{Simcoe:12} for their $z$=7.04 DLA (upper limits).  Note the observed equivalent width have been adjusted to the observed frame rather than rest-frame which was presented in the original work.   Direct qualitative comparisons should be made between $W_1$ and $W_{\rm obs}$.
}
\begin{tabular}{cccccl}
\hline
Ion & $\lambda_{\rm rest}$	& $W_1$  & $W_2$ & 	$W_3$	& $W_{\rm obs}$	\\  
	&  [\AA]			  		& [\AA]	 & [\AA]	&	[\AA] &	[\AA]		\\
\hline
\vspace{.05mm}\\
\ion{O}{i}	&	$1302.2$ 	 & $0.009$	& $0.084$	& $0.218$ & $\leq0.354$ \\
\ion{C}{ii}	&	$1334.5$ 	 & $0.033$	& $0.284$	& $0.616$ & $\leq0.137$\\
\ion{Si}{ii}&	$1260.4$ 	 & $0.008$  & $0.077$	& $0.200$ & $\leq0.072$\\
\ion{Fe}{ii}&	$2586.7$ 	 & $0.009$	& $0.091$	& $0.245$ & $\leq0.306$\\
\ion{C}{iv}	&	$1548.2$ 	 & $0.065$	& $0.511$	& $0.965$ & $\leq0.579$\\
\ion{Mg}{ii}&	$2796.4$ 	 & $0.141$	& $1.051$	& $1.889$ & $\leq0.555$\\
\ion{Fe}{ii}&	$2600.2$ 	 & $0.032$	& $0.248$	& $0.731$ & $\leq0.113$\\
\hline
\end{tabular}
\label{tbl:ew}
\end{center}
\end{table}

The abundance for each species is obtained by convolving estimates for the yields for zero-metallicity Type~II and PISN events \citep{Heger:02,Heger:10}, with the IMF employed in our simulations. By integrating our selected IMF (Equations~\ref{eq:mlarson} and \ref{eq:imf}) over the appropriate mass ranges for Type~II and PISNe (see Section~\ref{sec:imf}), we determine the mass in metals contributed by each event, which we then use to calculate a weighted mean of the metal yields for both event types from Table~\ref{tbl:lines}. The weights for each event are given by
\begin{equation}
w_{{\rm TypeII}}=\frac{Y_{{\rm TypeII}}m_{*,{\rm TypeII}}}{Y_{{\rm TypeII}}m_{*,{\rm TypeII}}+Y_{{\rm PISN}}m_{*,{\rm PISN}}}\approx0.13,
\end{equation}
\begin{equation}
w_{{\rm PISN}}=\frac{Y_{{\rm PISN}}m_{*,{\rm PISN}}}{Y_{{\rm TypeII}}m_{*,{\rm TypeII}}+Y_{{\rm PISN}}m_{*,{\rm PISN}}}\approx0.87.
\end{equation}
The resulting abundance pattern is thus directly coupled to the assumed IMF. Our final column density for each absorber then is
\begin{equation}
N_X = N_{\rm HI}Z\left(w_{\rm TypeII}X_{x,{\rm TypeII}}\mathcal{R}_x + w_{\rm PISN}X_{x,{\rm PISN}}\mathcal{R}_x \right)\, . 
\end{equation}
The values for $Y_x$ and $X_x$ can be found in Table~\ref{tbl:M/E} and \ref{tbl:lines}, respectively, and $\mathcal{R}_x$ is the mass ratio of the absorber and hydrogen ($m_x/m_{\rm H}$).

\begin{figure}
\begin{center}
\includegraphics[scale=0.24] {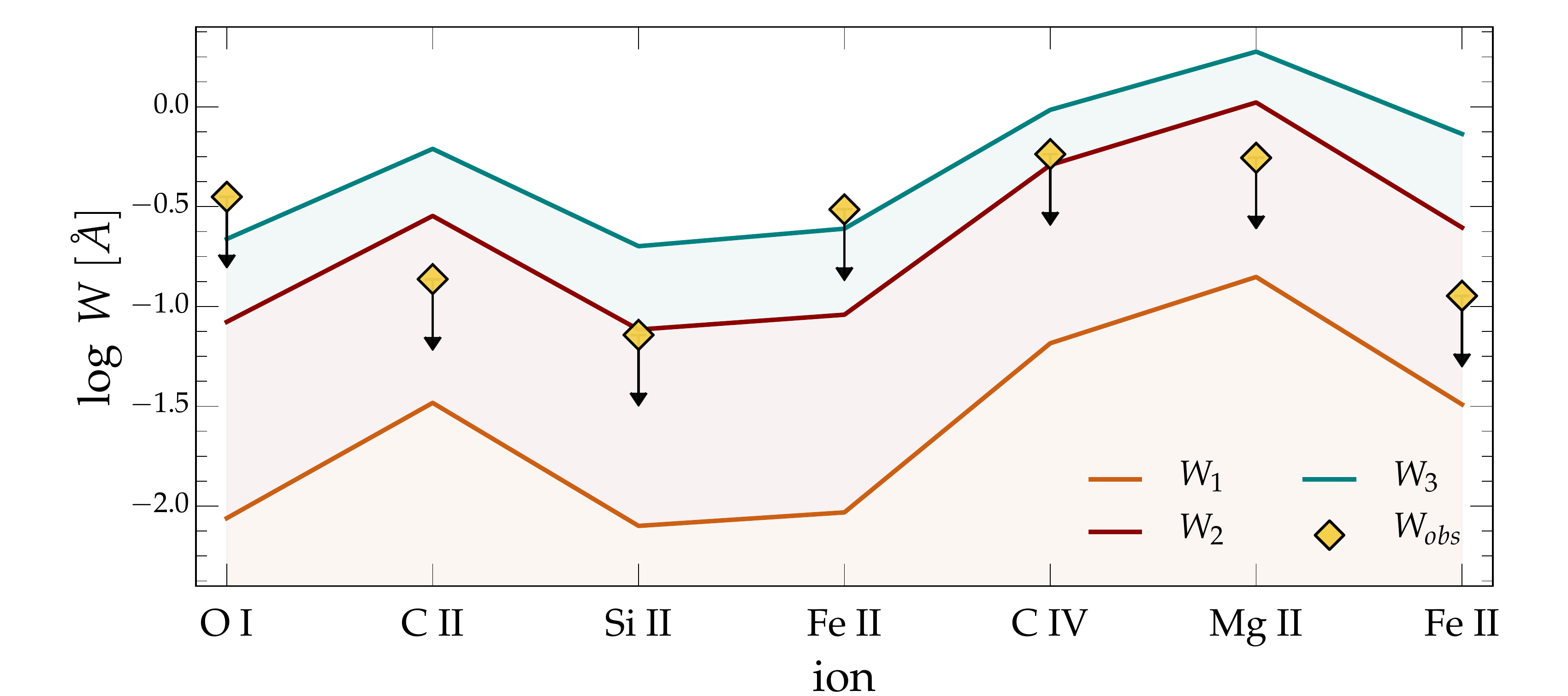}
\caption{Observer-frame equivalent widths, calculated from the synthetic spectra presented in Figure~\ref{fig:spec} and Table~\ref{tbl:ew}, compared to the $1\sigma$ upper limits derived from observations at $z$=7.04 from \citet{Simcoe:12} (yellow diamonds).  
} 
\label{fig:ew}
\end{center}
\end{figure}

In Figure~\ref{fig:spec}, we present select synthetic spectra, resulting from feature-less GRB afterglow emission which we pass through our simulated systems with properties similar to observed DLAs ($N_{\rm HI}\geq10^{20}\ {\rm cm^{-2}}$).  The first two spectra are take from regions which are similar in density and metallicity to the \citet{Simcoe:12} DLA estimates, and the final spectrum is randomly chosen from our parameter space in Figure~\ref{fig:colZ}.  While we make direct comparisons of our simulated equivalent widths to those found in observations, we caution the reader that our limited resolution and idealized nature of our methodology allow for only a qualitative comparison.

In Table~\ref{tbl:ew}, we provide estimates for observer-frame equivalent widths,
\begin{equation}
W=(1+z)\int[1-e^{-\tau(\lambda)}]\ d\lambda\, ,
\end{equation}
for select ions calculated from the spectra presented in Figure~\ref{fig:spec}.  The subscripts $1, 2, 3$ correspond to the spectra from top to bottom.  The final column, $W_{\rm obs}$, contains the values presented in \citet{Simcoe:12} for their $z$=7.04 DLA.  We note that the observed values were shifted to the observed frame for direct comparison to $W_{\rm obs}$. We find that our equivalent widths $W_1$ and $W_2$, which have similar column density ($\log N_{\rm HI}=20.6 {\rm cm^{-2}}$) and metallicity ($\log Z/\Zsun =-3, -4$) to the systems in \citet{Simcoe:12},  are in qualitative agreement with their published results.  This agreement suggests that a Pop~III only enriched region is a plausible explanation for the \citet{Simcoe:12} observation. 

We caution that due to the inherent difficulties in the observational estimates, and the large uncertainty in our adopted model abundances, any comparison should be viewed as a proof of concept only. 
In future work we intend to revisit this subject with a more sophisticated Pop~II star formation model enabled.

It is also worth noting that the volume filling fraction of systems with metallicities similar to \citet{Simcoe:12} ($Z/Z_\odot=10^{-4}-10^{-3}$) is $\mathcal{F}_V=10^{-5}-10^{-3}$, regardless of column density, suggesting that these systems are very rare in our simulation volume.


\section{Summary and Conclusions}
\label{sec:sum}
We develop a new sub-grid model for Pop~III star formation (P3L) in high-resolution cosmological volume simulations.  We utilize the P3L model to study the baseline metal enrichment of galaxies in the first billion years of cosmic history. We would like to remind the reader, when examining these results, that we are only considering metal enrichment from Pop~III star formation. Therefore, what we present here is an idealized numerical experiment, providing an upper limit to the ability of Pop~III to enrich the early Universe. The neglected supernova feedback from Pop~II star formation is expected to decrease the overall enrichment from Pop~III. Conversely, our simulations here establish the ceiling for Pop~III metal enrichment, which can in turn constrain attempts to interpret the chemical abundance record in extremely metal-poor objects. In future work, we will revisit our results with Pop~II star formation and feedback self-consistently included. Our major conclusions are as follows:

\begin{itemize} 
\item We find that our legacy Pop~III model of star formation produces a Pop~III SFRD which is consistent with both observations of Pop~II dominated systems, and numerical studies at $z\gtrsim7.5$. When the effects of the UV ionizing background are considered, the Pop~III SFRD peaks at a value of $\sim 10^{-3}\ {\rm \Msun yr^{-1} Mpc^{-1}}$ near $z\simeq10$.

\item The mean enrichment from Pop~III star formation rises smoothly between $z\simeq25-7$, but does not exceed the critical metallicity, $Z_{\rm crit}=10^{-4}\ \Zsun$, required for the Pop~III to Pop~II transition until $z\simeq7$.  

\item The baseline average metal enrichment from Pop~III star formation for bound, star forming systems (i.e. dark mater haloes) is $Z\sim 10^{-3}\ \Zsun$. We also find that a maximum enrichment (``Pop~III plateau``) of $Z\simeq 10^{-2}\ \Zsun$ emerges as star formation in haloes with $Z>Z_{\rm crit}$ must be fueled by the accretion of pristine gas creating a quasi equilibrium between mass growth and enrichment.

\item At $z$=7.5, {\it all} dark matter haloes with $M_h\geq 10^9\ \Msun$ contain metals from Pop~III star formation.  This fraction falls to 62\% of haloes with $M_h\geq 10^8\ \Msun$ and to 7\% for $M_h\geq 10^7\ \Msun$. 

\item Only a very small fraction ($\mathcal{F}_V\sim 10^{-2.8}$)of the simulation volume is enriched by Pop~III star formation to values beyond $Z_{\rm crit}$. For enrichment levels of $\log Z/\Zsun \geq -3, -2, -1$, the respective volume filling fractions are $\mathcal{F}_V\sim10^{-3}, 10^{-5}, 10^{-8}$ at $z$=7.5.

\item While locally star forming haloes are quickly enriched to $Z>Z_{\rm crit}$, globally Pop~III star formation is not a ``self-terminating'' process, and requires instead the influence of ionizing photons produced from Pop~II star formation. This is in contrast to conclusions from previous work \citep{Yoshida:04}. The primary difference in our study is the adoption of an IMF which allows for less massive Pop~III stars as opposed to only stars with $M_*\simeq100-300\ \Msun$ as in the \cite{Yoshida:04} work.  Our less top heavy IMF reduces both the metal yield and radius of enrichment for each Pop~III star formation event.

\item Within the bound systems we find that a bimodal distribution of haloes emerges which is distinguishable by either being internally or externally enriched.  The population of haloes which has been enriched by external processes (i.e. nearby halo star formation events, accretion of enriched gas, merger events) makes up approximately 45\% of the $z=7.5$ haloes.  These low mass systems exhibit a metallicity floor of $Z\simeq 10^{-6.5}\ \Zsun$, which is much lower than their more massive ($M_h\gtrsim10^{7.5}\ \Msun$), star-forming counterparts. By connecting progenitors and descendants, we find that the externally enriched systems found at $z$=20 quickly merge with the nearby star forming haloes.

\item Using absorption spectra created from a sample of our simulated volume which contains the unique abundance pattern of exclusive Pop~III enrichment, we find equivalent widths for selected ions in good agreement with $z$=7.04 DLA observations \citep{Simcoe:12}.  This agreement lends support to the authors' conclusion that the region could have been enriched by stars with a Pop~III-like IMF.

\item We find that, at $z$=7.5, the critical neutral hydrogen column density below which metals from Pop~III are no longer present is $N_{\rm HI,crit}\sim 10^{11} {\rm cm^{-2}}$.

\end{itemize}

Cosmic metal enrichment is a complex, multi-generational process, and it is intriguing to be able to directly probe the very first episodes of this long history. With ongoing simulations of high-$z$ star and galaxy formation, like the work presented here, we are constructing a heuristic net for discovery with the upcoming frontier telescopes, such as the {\it JWST} and the extremely-large telescopes on the ground (GMT, TMT, E-ELT).




\section*{Acknowledgments}
Support for Program number HST-AR-14569.001-A (PI Jaacks) was provided by NASA through a grant from the Space Telescope Science Institute, which is operated by the Association of Universities for Research in Astronomy, Incorporated, under NASA contract NAS5-26555.  VB is supported by NSF grant AST-1413501.  This work used the Extreme Science and Engineering Discovery Environment (XSEDE), which is supported by National Science Foundation grant number ACI-1053575, allocation number TG-AST120024.


\bibliographystyle{mnras}

\end{document}